\documentclass[fleqn,usenatbib]{mnras}

\usepackage{newtxtext,newtxmath}

\usepackage[T1]{fontenc}
\usepackage{ae,aecompl}

\usepackage{graphicx}	
\usepackage{amsmath}	
\usepackage[dvipsnames]{xcolor}
\usepackage{multirow}
\usepackage{delarray}
\usepackage{color}
\usepackage{here}
\usepackage{float}
\usepackage{tensor}
\usepackage{xspace}
\usepackage{orcidlink}
\usepackage[multiple]{footmisc}

\newcommand{\mach}{\mathcal{M}}

\newcommand{\be}{\begin{equation}} \newcommand{\ee}{\end{equation}}

\newcommand{\solarmass}{\mathrm{M}_{\rm \sun}}
\newcommand{\msun}{\solarmass}
\newcommand{\Mmedian}{M_{\rm 50}}
\newcommand{\Msonic}{M_{\rm sonic}}
\newcommand{\MBE}{M_{\rm BE}^{\rm turb}}

\newcommand{\Lsonic}{L_{\rm sonic}}
\newcommand{\MJeans}{M_{\rm Jeans}}
\newcommand{\alphath}{\alpha_{\mathrm{th}}}
\newcommand{\alphaturb}{\alpha_{\mathrm{turb}}}
\newcommand{\alphaturbzero}{\alpha_{\mathrm{turb,0}}}
\newcommand{\alphaB}{\alpha_{\mathrm{B}}}
\newcommand{\cs}{c_{\rm s}}

\newcommand{\pc}{\mathrm{pc}}
\newcommand{\kelvin}{\mathrm{K}}

\newcommand{\dderiv}{\mathrm{d}}
\newcommand{\vvector}{\mathbf{v}}
\newcommand{\Bvector}{\mathbf{B}}
\newcommand{\appropto}{\mathrel{\vcenter{
  \offinterlineskip\halign{\hfil$##$\cr
    \propto\cr\noalign{\kern2pt}\sim\cr\noalign{\kern-2pt}}}}}

\newcommand{\myquote}[1]{``#1''}

\usepackage[normalem]{ulem}

\title[Can magnetized turbulence set the mass scale of stars?]{Can magnetized turbulence set the mass scale of stars?}

\author[]{
D\'avid Guszejnov\orcidlink{0000-0001-5541-3150}$^{1}$\thanks{guszejnov@utexas.edu},
Michael Y. Grudi\'{c}\orcidlink{0000-0002-1655-5604}$^{2}$\thanks{mike.grudic@northwestern.edu},
Philip F. Hopkins\orcidlink{0000-0003-3729-1684}$^{3}$,
\newauthor
Stella S. R. Offner\orcidlink{0000-0003-1252-9916}$^{1}$,
Claude-Andr{\'e} Faucher-Gigu{\`e}re\orcidlink{0000-0002-4900-6628}$^{2}$
\\
$^{1}$Department of Astronomy, University of Texas at Austin, TX 78712, USA \\
$^{2}${CIERA and Department of Physics and Astronomy, Northwestern University, 2145 Sheridan Road, Evanston, IL 60208, USA}\\
$^{3}$TAPIR, Mailcode 350-17, California Institute of Technology, Pasadena, CA 91125, USA \\
}

\date{\today \vspace{-0.6cm}}

\pubyear{2019}

\begin{document}
\label{firstpage}
\pagerange{\pageref{firstpage}--\pageref{lastpage}}
\maketitle

\begin{abstract}
Understanding the evolution of self-gravitating, isothermal, magnetized gas is crucial for star formation, as these physical processes have been postulated to set the initial mass function (IMF). We present a suite of isothermal magnetohydrodynamic (MHD) simulations using the {\small GIZMO} code, that follow the formation of individual stars in giant molecular clouds (GMCs), spanning a range of Mach numbers found in observed GMCs ($\mathcal{M} \sim 10-50$). As in past works, the mean and median stellar masses are sensitive to numerical resolution, because they are sensitive to low-mass stars that contribute a vanishing fraction of the overall stellar mass. The {\em mass-weighted} median stellar mass $M_\mathrm{50}$ becomes insensitive to resolution once turbulent fragmentation is well-resolved. Without imposing Larson-like scaling laws, our simulations find  $M_\mathrm{50} \appropto M_\mathrm{0} \mach^{-3} \alpha_\mathrm{turb} \mathrm{SFE}^{1/3}$ for GMC mass $M_\mathrm{0}$, sonic Mach number $\mathcal{M}$, virial parameter $\alpha_\mathrm{turb}$, and star formation efficiency $\mathrm{SFE}=M_\mathrm{\star}/M_\mathrm{0}$. This fit agrees well with previous IMF results from the {\small RAMSES}, {\small ORION2}, and {\small SphNG} codes. Although $\Mmedian$ has no significant dependence on the magnetic field strength at the cloud scale, MHD is necessary to prevent a fragmentation cascade that results in non-convergent stellar masses. For initial conditions and SFE similar to star-forming GMCs in our Galaxy, we predict $\Mmedian$ to be $>20 M_{\odot}$, an order of magnitude larger than observed ($\sim 2 M_\odot$), together with an excess of brown dwarfs. Moreover, $\Mmedian$ is sensitive to initial cloud properties and evolves strongly in time within a given cloud, predicting much larger IMF variations than are observationally allowed. We conclude that physics beyond MHD turbulence and gravity are necessary ingredients for the IMF.
\end{abstract}



 \section{Introduction}\label{sec:intro}
 Star formation involves many physical mechanisms acting in concert, including gravity, hydrodynamics, magnetic fields, radiation and chemistry. While all of these processes have a role to play, understanding the whole picture is difficult without first understanding how various subsets of these mechanisms work together. Above all, it is important to explore how star formation arises from the interplay of gravity and turbulence, which provide the canvas upon which other physics can be painted.
 
 The simplest and best-studied model of star formation considers only the equations of isothermal hydrodynamics coupled to gravity, which models the dense, $\sim 10\rm K$ interstellar medium (ISM) found in molecular clouds in our Galaxy \citep[e.g.,][]{padoan_nordlund_2002_imf,hc08,core_imf}. Many numerical works studying star formation in turbulent molecular clouds in this framework have found the problem to be ill-posed: numerical convergence in the mass spectrum of collapsed fragments, which should map onto the stellar Initial Mass Function (IMF), is typically not achieved \citep[see e.g.][]{Martel_numerical_sim_convergence,Kratter10a,Federrath_2017_IMF_converge_proceedings, guszejnov_isothermal_collapse, Hennebelle_Lee_isoT_sim}. \citet{larson2005} noted that an isothermal, self-gravitating medium can spontaneously form filamentary structures that formally collapse to infinite density before they break apart \citep[e.g.][]{truelove_1997_dens_condition}, so that the collapsed mass cannot be meaningfully discretized into individually-collapsing cores, as predicted analytically by \citet{Inutsuka_Miyama_1992} for an idealized filament. Even if cores do form, they can sub-fragment indefinitely in a self-similar fashion (see \citealt{guszejnov_feedback_necessity,guszejnov_isothermal_collapse}, for a counter-argument see \citealt{Andre_2019_filaments_IMF}). Thus it is not clear that isothermal gas physics and gravity alone can meaningfully predict {\it any} IMF, let alone the observed one. 
 
However, molecular clouds are observed to have a non-negligible amount of magnetic support 
\citep{crutcher_2009_mc_magnetic_fields}. The introduction of magnetic fields can suppress the growth of the Jeans instability \citep{chandra_fermi_mhd_jeans}, support structures against collapse \citep{Mouschovias_Spitzer_1976_magnetic_collapse}, and cushion supersonic shocks that may form dense structures, generally reducing the rate of star formation and the degree of fragmentation in molecular clouds (e.g. \citealt{price_bate_2008_mhd_cluster,federrath_2015_inefficient_sf}, see \citealt{Krumholz_2019_IMF_magnetic_field} and \citealt{Hennebelle_2019_MHD_cloud_evol} for reviews). Due to their ability to suppress fragmentation, magnetic fields have long been considered potential candidates for setting the mass scales of stars \citep[e.g.,][]{Shu_1987_star_formation, Mckee_tan_2003_turbulent_core, padoan_nordlund_2011_imf}. But similar to the non-magnetized case, the ideal magnetohydrodynamic (MHD) equations governing the evolution of the gas have no inherent physical scale \citep{sf_big_problems} of their own, so any mass scale in stellar masses must be imposed by initial and boundary conditions. In the non-magnetized case the initial conditions are washed out by a turbulent fragmentation cascade, ultimately imposing no physical mass scale in the IMF \citep{guszejnov_isothermal_collapse}. For magnetized gas, recent high resolution simulations have claimed convergence \citep[e.g.,][]{Haugbolle_Padoan_isot_IMF} in the mass function (or more specifically, that the mass spectrum of sink particles is insensitive to numerical resolution), while other works with similar numerical resolutions have argued for non-convergence \citep[i.e.\ strong resolution-dependence][]{Federrath_2017_IMF_converge_proceedings}. 

In this paper we use numerical MHD simulations, achieving a dynamic range in mass resolution an order of magnitude higher than any previous star cluster formation studies and covering a broad parameter space (see \S~\ref{sec:results}), to explore the following questions: Is there a characteristic mass in the initial conditions of ideal isothermal MHD that is inherited by the mass function of the final fragments? How does this characteristic mass depend on initial conditions, such as the sonic and Alfvén Mach numbers? Could this characteristic mass set the mass scale of stars? Note that the original algorithm used in the paper had a bug in the sink particle algorithm, leading to an excess of very-low-mass objects. This does not change the results and is addressed in detail in the erratum in Appendix \ref{sec:erratum}.


 



 \section{Methods}
 
 \subsection{Ideal isothermal MHD}
 
 \subsubsection{MHD equations}

An isothermal, magnetized, infinitely conducting, self-gravitating fluid (well above the dissipation scale) is completely described by the following closed set of dimensionless equations (see \citealt{Mckee_2010_MHD} for a more detailed derivation):
\begin{eqnarray}
\label{eq:mom_eq_dimless}
\frac{\partial }{\partial \tilde{t}}\left(\tilde{\rho}\right)+\tilde{\nabla}\cdot\left(\tilde{\rho}\tilde{\vvector}\right)=0,\nonumber\\
\frac{\partial }{\partial \tilde{t}}\left(\tilde{\rho} \tilde{\vvector}\right)+\tilde{\nabla}\cdot\left(\tilde{\rho}\tilde{\vvector}\otimes\tilde{\vvector}\right)=
-\tilde{\nabla} \tilde{\rho}-\frac{15}{4 \pi}\alphath^{-1}\tilde{\rho}\tilde{\nabla}\tilde{\Phi}-2\beta^{-1}\left(\tilde{\nabla} \times \tilde{\Bvector}\right)\times \tilde{\Bvector},\nonumber \\
\tilde{\nabla}^2\tilde{\Phi}=4\pi \tilde{\rho},\nonumber \\
\frac{\partial }{\partial \tilde{t}}\tilde{\Bvector} + \tilde{\nabla} \times \left(\tilde{\Bvector} \times \tilde{\vvector}\right) = 0,
\end{eqnarray}
where $\tilde{\rho}\equiv \rho/\rho_0$, $\tilde{\vvector} \equiv {\vvector}/{\cs}$, $\tilde{t} \equiv t\,\cs/L_{0}$, $\tilde{\nabla} \equiv L_{0}\,\nabla$, and $\tilde{\Bvector}=\Bvector/B_0$ are the normalized fluid density, velocity, time, gradient, and the magnetic field, $\cs=\rm const.$ is the isothermal sound speed and $\tilde{\Phi}\equiv\frac{\Phi}{G \rho_0 L_0^2}$ is the dimensionless gravitational potential. Meanwhile, $\alphath\equiv \frac{15}{4 \pi}\frac{\cs^2}{G \rho_0 L_0^2}$ is the (thermal) virial parameter, which is equivalent to the ratio of thermal to gravitational energy in a homogeneous sphere of radius $L_0$. Meanwhile, $\beta \equiv P_{\mathrm{thermal,0}}/P_{\mathrm{magnetic,0}} = 2\cs^2/v_{A,\,0}^2$ is the characteristic plasma beta, where $P_{\mathrm{thermal,0}}$, $P_{\mathrm{magnetic,0}}$ are the characteristic thermal and magnetic pressures of the system respectively, while $v_{A,\,0}^{2} \equiv B_{0}^{2}/(\mu_0\rho_{0})$ is the Alfv\'{e}n speed of the fluid at $B_{0}$ and $\rho_{0}$ with $\mu_0$ being the vacuum permeability. It is also useful to introduce the 3D sonic Mach number $\mach^2\equiv\langle ||\vvector||^2/\cs^2\rangle=\langle ||\tilde{\vvector}||^2\rangle$. 


Note that as defined above, $\rho_{0}$, $\cs$, $B_{0}$, and $L_{0}$ are simply arbitrary normalization units: for convenience in our study here, we will take these to be the mean initial values of the clouds studied (giving the usual meaning to the virial parameter, $\beta$, and Mach number, in a cloud-averaged sense). With these definitions, the thermal virial parameter $\alphath$, the plasma $\beta$ and the Mach number $\mach$ each describe the relative weight of the different processes in the momentum equation (and are defined by mean cloud properties in the initial conditions). In other words, the dynamics are \emph{entirely} determined by the three dimensionless constants $\alphath$, $\beta$ and $\mach$, for a given initial condition. The only way to impose a characteristic scale on the problem (such as a characteristic mass for collapsing cores) is through these initial conditions.

\subsubsection{Parameters and mass scales}\label{sec:params_scales}

Here we summarize the main mass scales and physical parameters that can be derived from the initial conditions, which will inform our analysis of the characteristic scales/mass relationships discussed in \S~\ref{sec:results}.

Due to the dimensionless nature of the system (see Eq. \ref{eq:mom_eq_dimless}), all mass scales must be inherited from initial conditions and their relative magnitude is described by $\alphath$, $\beta$ and $\mach$. In the literature it is common to introduce alternate parameters, like the \emph{turbulent virial parameter} (see \citealt{Bertoldi_McKee_1992}):
\be
\alphaturb \equiv  \frac{ 2 E_{\rm turb}}{-E_{\rm grav}} = \alphath \frac{1}{3}\mach^2,
\label{eq:alpha_turb}
\ee
the \emph{magnetic virial parameter}:
\be
\alphaB \equiv \frac{2 E_{\rm mag}}{-E_{\rm grav}} = \frac{2\alphath}{3 \beta},
\ee
and the \emph{total virial parameter}:
\be
\alpha \equiv 2\frac{ E_{\rm th} + E_{\rm turb}+ E_{\rm rot} +  E_{\rm mag}}{-E_{\rm grav}} = \alphath\left[1+\frac{1}{3}\left(\mach^2+\mach_{\mathrm{rot}}^2+\frac{2}{\beta}\right)\right],
\ee
where $E_{\rm th}$, $E_{\rm turb}$, $E_{\rm rot}$, $E_{\rm mag}$ and $E_{\rm grav}$ are the turbulent kinetic, rotational, thermal, magnetic and gravitational binding energies of the gas, while $\mach_{\mathrm{rot}}\equiv v_{\mathrm{rot}}/\cs$ and $v_{\mathrm{rot}}$ is the average rotational velocity within the system.

Thermal pressure can prevent the collapse of a fluid element, where the corresponding mass scale (up to arbitrary order-unity constants) is the \emph{Jeans mass}:
\be
\frac{\MJeans}{M_0} \equiv \frac{\frac{4 \pi}{3}\rho_0 \left(\frac{\cs}{\sqrt{G \rho_0}}\right)^3}{\frac{4 \pi}{3}\rho_0 L_0^3} = \left(\frac{4\pi}{15} \alphath\right)^{3/2}= \left(\frac{4\pi}{5}\right)^{3/2}\alphaturb^{3/2}\mach^{-3}.
\label{eq:MJeans}
\ee
Note that we normalize the Jeans mass and other mass scales below in units of $M_{0} \equiv 4\pi \rho_{0}\,L_{0}^{3}/3$, the characteristic mass scale (e.g.\ total cloud mass in a spherical cloud), so that we can write it only in terms of the key dimensionless parameters above. The initial turbulence also has a characteristic length scale: the sonic length, $\Lsonic$, on which the turbulent dispersion becomes supersonic. The corresponding mass scale is the \emph{sonic mass}:
\be
\frac{\Msonic}{M_0} \equiv \frac{\cs^2 \Lsonic}{G \rho_0 L_0^3}=\frac{4\pi}{15}\alphath \mach^{-2}= \frac{4\pi}{5}\alphaturb\mach^{-4},
\label{eq:Msonic}
\ee
where we used the supersonic linewidth-size relation ($\sigma^2(L)\propto L$). Another mass scale of an isothermal turbulent flow is the \textit{turbulent Bonnor-Ebert mass}, the maximum gas mass that can support itself against its own self-gravity plus external pressure in post-shock compressed gas with $\tilde{\rho} \sim 1+\frac{1}{3}\mach^2$  \citep{padoan_nordlund_1997_imf}, which scales as
\be
\frac{\MBE}{M_0} \sim 2 \frac{\MJeans}{M_0}\left(1+\frac{1}{3}\mach^2\right)^{-1/2}=  \frac{2 \left(\frac{4\pi}{5}\right)^{3/2}\alphaturb^{3/2}}{\left(1+\frac{1}{3}\mach^2\right)^{1/2}\mach^{3}}.
\label{eq:MBE}
\ee

The initial magnetic field can also impose a mass scale, below which magnetic fields provide enough support to prevent collapse \citep{Mouschovias_Spitzer_1976_magnetic_collapse}. This relative \emph{ magnetic critical mass} is:
\be
\frac{M_{\Phi}}{M_0}\equiv \sqrt{\alphaB} \sim \sqrt{\frac{2\alphath}{3\beta}}.
\label{eq:M_Phi}
\ee
It is common to introduce a very similar measure, the \emph{normalized magnetic flux} (or mass-to-flux ratio):
\be
\mu \equiv c_1 \sqrt{2} \frac{M_0}{M_{\Phi}}= c_1\sqrt{\frac{2}{\alphaB}} \sim c_1\sqrt{\frac{3\beta}{\alphath}},
\ee
where $c_1\approx 0.4$. With this normalization $\mu=1$ corresponds to the critical point in the stability of a homogeneous sphere in a uniform magnetic field \citep{Mouschovias_Spitzer_1976_magnetic_collapse}. 

Due to their prevalence in the literature, we describe our runs with the dimensionless parameters $\alphaturb$, $\mu$ and $\mach$ (which are mathematically equivalent to $\alpha_{\rm th}$, $\beta$, and $\mach$) in the remainder of this paper.

\subsection{Simulations}
\subsubsection{Numerical methods}
  Here we briefly summarize our numerical approach to simulating star-forming GMCs, but defer a full description and presentation of numerical tests to an upcoming methods paper (Grudi\'{c} et al. 2020, in prep.). Similar to our study of non-magnetized isothermal collapse \citep{guszejnov_isothermal_collapse}, we simulate star-forming clouds with the {\small GIZMO} code\footnote{\url{http://www.tapir.caltech.edu/~phopkins/Site/GIZMO.html}} (\citealt{hopkins_mfm_2015}), using the Lagrangian meshless finite-mass (MFM) method for magnetohydrodynamics \citep{hopkins_gizmo_mhd}, with numerous upgrades and optimizations to make the code suitable for simulating star formation and stellar dynamics, including a new set of timestep criteria based on \citet{grudic_tidal_timestep}. We use the \citet{hopkins_mhd_cg} constrained-gradient scheme to ensure the $\nabla \cdot \mathbf{B}=0$ constraint is satisfied to high precision. The gas obeys an isothermal equation of state with $c_s=0.2\,\rm km/s$ (effective gas temperature $T\sim10\,\kelvin$)
  in our adopted code units, however the equations solved are scale-free, so this choice of $c_s$ is arbitrary. Gravity is solved with the approximate Barnes-Hut tree method \citep{Springel_2005_gadget}. Force softening is fully adaptive for gas cells \citep{price_monaghan_softening, hopkins2015_gizmo}, with no imposed floor. Sink particles (representing stars) have a fixed Plummer-equivalent softening radius of $7.56\,\rm AU$, unlike \citet{guszejnov_isothermal_collapse} where we also used adaptive softening for sink particles. As such we are able to follow the formation and evolution of binaries and multiples with separations larger than $\sim 10\,\rm AU$.
  
  To carry on the calculation past the runaway collapse of the first core, we use a sink particle algorithm very similar to \citet{Bate_1995_accretion}. A gas cell is converted to a sink particle if it satisfies a number of criteria intended to identify the centres of collapsing cores that have become too dense to resolve the Jeans instability \citep{Bate_1995_accretion,truelove_1997_dens_condition, Federrath_2010_sinks,gong_2013_athena_sinks}. We take this density threshold to be
  \begin{equation}
      \rho_{\rm J} = \frac{\pi^3 c_{\rm s}^6}{64 G^3 \Delta m^2} = 3 \times 10^{-14}\mathrm{g\,cm}^{-3} \left(\frac{c_s}{0.2\,\mathrm{km\,s}^{-1}}\right)^6\left(\frac{\Delta m}{10^{-3} \msun}\right)^{-2}
      \label{eq:rhoJ}
  \end{equation}
  (where $\Delta m$ is the conserved cell mass) 
  corresponding to the density at which a hydrodynamic cell of size $\Delta x = \left(\Delta m/\rho\right)^{1/3}$ contains half a Jeans wavelength $\lambda_J=c_s\sqrt{\frac{\pi}{G \rho}}$. Cells converted to sinks must also be a local density maximum among their $N_\mathrm{ngb} \sim 32$ nearest neighbors, be gravitationally bound accounting for thermal, turbulent, and magnetic energy \citep{Federrath_2010_sinks, hopkins2013_sf_criterion}, and must be collapsing along all 3 axes \citep{gong_2013_athena_sinks}. Lastly, we impose a new tidal criterion to be described fully in Grudi\'{c} et al. 2020 (in prep.) that is similar in motivation to the potential-minimum criterion of \citet{Federrath_2010_sinks}, but is invariant to the transformation $\mathbf{g}\rightarrow \mathbf{g}+\mathbf{g}'$, where $\mathbf{g}'$ is a constant, uniform acceleration that should have no effect upon the system's internal dynamics \citep[see][]{bleuler_teyssier_sinks}.
  
  Sink particles interact with gas cells via gravity and accretion. To be accreted by a sink, gas cells must lie within the sink radius
  \begin{equation}
      R_{\rm sink} = \max \left(\left(\frac{3\Delta m}{4 \pi \rho_{\rm J}}\right)^{1/3}, 21\rm AU \right),
  \end{equation}
  the greater of the volume-equivalent spherical radius of a gas cell of density $\rho_{\rm J}$ or the support radius of the sink's gravitational softening kernel (ie. $2.8 \times 7.56\rm AU$). To be accreted, cells must also be gravitationally bound to the sink and must have less angular momentum than a circular orbit at $R_{\rm sink}$. When a gas cell is accreted, its mass, momentum, center of mass moment, angular momentum, and magnetic flux are transferred to the sink particle. This is essentially the same prescription that other Lagrangian codes use (e.g., \citealt{price_bate_2007_mhd_sf,Price_2012_Sph_MHD,Wurster_2019_no_magnetic_break_catastrophe}). The accreted angular momentum is redistributed to nearby gas cells with an e-folding time equal to the freefall time at $\rho_{\rm J}$, similar to the prescription of \citealt{hubber:2013.sinks}. Note that we have experimented with several variations to the above prescriptions, including using different values for the critical density relative to $\rho_{\rm J}$, varying the sink radius and removing magnetic energy from the boundedness condition. We will present the results of these experiments in detail in a future numerics-focused work (Grudi\'{c} et al. in prep.), but can summarize that none of the results in the present work are sensitive to these choices.

\subsubsection{Initial conditions}
 
For the runs included in this paper we are using two different sets of initial conditions (ICs) common in the literature, to ensure that our results are robust to the specifics of the IC generation\footnote{The initial conditions are generated by the \href{https://github.com/mikegrudic/MakeCloud}{{\small MakeCloud}} script.}:
\begin{itemize}
\item {\bf Sphere ICs} begin with a spherical cloud ($T=10\,\kelvin$, the radius $R_\mathrm{cloud}$ and mass $M_\mathrm{0}$ are specified) with uniform density, surrounded by diffuse gas with a density contrast of 1/1000. The cloud is placed at the center of a $10 R_\mathrm{cloud}$ box, that is periodic to gas cells and sink particles but not for gravitational forces (has no discernible effect, but reduces computational cost). The velocity field is 
a Gaussian random field with power spectrum $E_k\propto k^{-2}$ \citep{ostriker_2001_mhd}, generated on a Cartesian grid and interpolated to the cell positions. The magnitude of the velocity field is rescaled to the value prescribed by $\alpha_{\rm turb}$. The initial clouds have a uniform $B_z$ magnetic field whose strength is set by the parameter $\mu$. There is no external driving in these simulations. Note that for these simulations, we define $\alpha_\mathrm{turb}$ similar to how previous studies did in the literature (e.g., \citealt{Bertoldi_McKee_1992, federrath_sim_2012}),
\be
\alphaturbzero \equiv \frac{5 c_s^2 \mach^2 R_\mathrm{cloud}}{3 G M_\mathrm{0}}.
\label{eq:alphaturb_sphere}
\ee
Note that this matches the definition from Eq. \ref{eq:alpha_turb} for a spherical cloud, so $\alphaturbzero=\alphaturb$ in these cases, but can significantly differ for different initial conditions \citep[see][]{federrath_sim_2012}. Nevertheless, it is a parameter that describes the relative importance of the initial turbulence to gravity.

\item {\bf Box ICs} are initialized with the cells set up on a uniform 3D grid, each starting at zero velocity and $T=10\,\kelvin$. The boundary conditions of this box are periodic for both hydrodynamics and gravity. This periodic box is then \myquote{stirred} by running the simulation with a pre-determined turbulent driving spectrum ($E_k\propto k^{-2}$, i.e. supersonic turbulence) and an appropriate decay time for driving mode correlations ($t_{\mathrm{decay}}\sim t_{\mathrm{cross}}\sim L_{\rm box}/\sigma_{\rm 3D}$) \citep{federrath_sim_compare_2010,bauerspringel2012}. This stirring is initially performed without gravity for 5 global freefall times $\left(t_{\mathrm{ff}}\equiv\sqrt{\frac{3 \pi}{32 G \rho_0}}\right)$. The result is a state of saturated MHD turbulence in which the density distribution is roughly log-normal, and correlations between the density, velocity, and magnetic fields are representative of realistic MHD turbulence. The normalization of the driving spectrum is set so that in equilibrium the gas in the box has a turbulent velocity dispersion ($\sigma_{\rm 3D}$) that gives the desired $\mach$ and $\alpha_{\rm turb}$. We use purely solenoidal driving, which remains active throughout the simulation after gravity is switched on (see Section \ref{sec:box_vs_sphere} for a discussion on this choice).  We take the box side length $L_\mathrm{box}$ to give a box of equal volume to the associated {\it Sphere} cloud model, i.e. $L_\mathrm{box} = \left(\frac{4\uppi}{3}\right)^{1/3}R_\mathrm{cloud}$, and thus define $\alphaturbzero$ using the volume-equivalent $R_\mathrm{cloud}$ in Equation \ref{eq:alphaturb_sphere}.
\end{itemize}

Table \ref{tab:IC} shows the target parameters for the runs we present in this paper. The input parameters are the turbulent virial parameter $\alphaturbzero$, normalized magnetic flux $\mu$ and Mach number $\mach$, which, together, fully define the initial conditions due to the scale-free nature of the problem. Using the mass-size relation of observed GMCs in the Milky Way (e.g. \citealt{larson_law}, specifically assuming $\Sigma\equiv M_\mathrm{0}/ \uppi R_\mathrm{cloud}^2 = 63 \msun\,\mathrm{pc}^{-2}$) we can identify the observable counterparts of these clouds, which are molecular clouds between $2000$ - $2\times 10^6\,\msun$. For each set of parameters in Table \ref{tab:IC} we carried out both \textit{Sphere} and \textit{Box} runs at several resolution levels. An important difference between the \textit{Sphere} and \textit{Box} runs is that in case of driven boxes the magnetic field is enhanced by a turbulent dynamo \citep{federrath_2014_dynamo} and saturates at about $\alphaB\sim 0.1$. This means that: 1) for Box runs $\mu$ is not a free parameter and 2) by doing both kinds of runs we are effectively exploring the effects of changing $\mu$. Note that of the $\alphaturbzero$, $\mu$, $\mach$ parameter space we concentrate on the region relevant to the description of star forming GMCs in the present-day Milky Way (outside of the galactic center). These clouds are highly supersonic ($\mach\gg 1$), have finite, but low magnetic support ($\mu > 1$) and negligible rotation aside from turbulent motions ($\mach_{\mathrm{rot}}=0$), see \citet{heyer_dame_2015} for a review. In this regime we can simplify Equations \ref{eq:alpha_turb}-\ref{eq:MBE} as approximately
\begin{eqnarray}
\alpha \approx \alphaturb = \frac{1}{3}\alphath \mach^{2},\label{eq:alpha_equiv}\\
\frac{\MJeans}{M_0} \approx  4\,\alphaturb^{3/2} \mach^{-3},\label{eq:mjeans_simple}\\
\frac{\Msonic}{M_0} \approx 2.5\,  \alphaturb \mach^{-4}\label{eq:msonic_simple}\\
\frac{\MBE}{M_0} \approx 14\,  \alphaturb^{3/2} \mach^{-4}\label{eq:MBE_simple}.
\end{eqnarray}
Since most Milky Way (MW) GMCs achieve a star formation efficiency ($\mathrm{SFE}=M_{\star}/M_{0}$) of 1\%-10\% over their lifetime (see \citealt{sf_big_problems} for a discussion, and note that some clouds have <1\%, see \citealt{Federrath_density_distrib}), we restrict our analysis to the SFE<10\% range, even though all of our simulations eventually reach $\mathrm{SFE}\sim 1$.

\begin{table*}
    \setlength\tabcolsep{2.0pt} 
	\centering
		\begin{tabular}{|cccc|cccc|cccccccc|c|}
		 \multicolumn{1}{c}{}&
		 \multicolumn{3}{c}{\bf Input Parameters} &
		 \multicolumn{4}{c}{\bf Scaled Parameters}&
		 \multicolumn{8}{c}{\bf Derived Parameters}& \multicolumn{1}{c}{\textbf{Resolution}} \\
		\hline
		\bf Key & $\alpha_{\rm turb}$ & $\mu$ & $\mach$ & $M_0$ [$\msun$] & $L_{\mathrm{box}}$ [pc] & $R_{\mathrm{cloud}}$ [pc] & $\cs$ [m/s] & $\alphath$ & $\alpha$ & $\mach_{\rm A} $ & $\beta$ & $\alpha_{\rm B}$ & $\frac{\MJeans}{M_0}$ & $\frac{\Msonic}{M_0}$ & $\frac{M_{\Phi}}{M_0}$ &  $\mathrm{max}\left(M_0/\Delta m\right)$ \\
		\hline		
		\bf M2e3\_R3 & 2 & 4.2 & 9.3 & 
		$2\times 10^3$ & 4.8 & 3 & 200 & 
		0.02 & 2.02 & 10 &  2.3 & 0.02 & 
		$1\times 10^{-2}$ & $6 \times 10^{-4}$ & 0.1 &  $10^{8}$ \\
		\hline
		\bf M2e4\_R10 & 2 & 4.2 & 16 & 
		$2\times 10^4$ & 16 & 10 & 200 & 
		0.008 & 2.02 & 10 &  0.78 & 0.02 & 
		$3\times 10^{-3}$ & $7 \times 10^{-5}$ & 0.1 &  $2\times10^{8}$ \\
		\hline
		\bf M2e5\_R30 & 2 & 4.2 & 29 & 
		$2\times 10^5$ & 48 & 30 & 200 & 
		0.002 & 2.02 & 10 &  0.23 & 0.02 & 
		$5\times 10^{-4}$ & $7 \times 10^{-6}$ & 0.1 &  $2\times10^{8}$ \\
		\hline
		\bf M2e6\_R100 & 2 & 4.2 & 51 & 
		$2\times 10^6$ & 160 & 100 & 200 & 
		0.0008 & 2.02 & 10 &  0.078 & 0.02 & 
		$8\times 10^{-5}$ & $7 \times 10^{-7}$ & 0.1 &  $2\times10^{8}$ \\
		\hline
		\bf M2e4\_R20\_a4 & 4 & 4.2 & 16 & 
		$2\times 10^4$ & 32 & 20 & 200 & 
		0.016 & 4.02 & 14 &  1.6 & 0.02 & 
		$8\times 10^{-3}$ & $2 \times 10^{-4}$ & 0.1 &  $2\times10^{7}$ \\
		\hline
		\bf M2e4\_R5\_a1 & 1 & 4.2 & 16 & 
		$2\times 10^4$ & 8 & 5 & 200 & 
		0.0039 & 1.02 & 7 &  0.39 & 0.02 & 
		$1\times 10^{-3}$ & $4 \times 10^{-5}$ & 0.1 &  $2\times10^{7}$ \\
		\hline
		\bf M2e4\_R2.5\_a0.5 & 0.5 & 4.2 & 16 & 
		$2\times 10^4$ & 4 & 2.5 & 200 & 
		0.002 & 0.52 & 5 &  0.19 & 0.02 & 
		$3\times 10^{-4}$ & $2 \times 10^{-5}$ & 0.1 &  $2\times10^{7}$ \\
		\hline
		\bf M2e4\_R1.25\_a0.25 & 0.25 & 4.2 & 16 & 
		$2\times 10^4$ & 2 & 1.25 & 200 & 
		0.0001 & 0.27 & 3.5 &  0.097 & 0.02 & 
		$1\times 10^{-4}$ & $1 \times 10^{-5}$ & 0.1 &  $2\times10^{7}$ \\
		\hline
		\bf M2e4\_R10\_mu13 & 2 & 13.4 & 16 & 
		$2\times 10^4$ & 16 & 10 & 200 & 
		0.008 & 2.002 & 31 &  7.8 & 0.002 & 
		$3\times 10^{-3}$ & $7 \times 10^{-5}$ & 0.04 &  $2\times10^{7}$ \\
		\hline
		\bf M2e4\_R10\_mu1.3 & 2 & 1.34 & 16 & 
		$2\times 10^4$ & 16 & 10 & 200 & 
		0.008 & 2.2 & 3.1 &  0.078 & 0.2 & 
		$3\times 10^{-3}$ & $7 \times 10^{-5}$ & 0.4 &  $2\times10^{7}$ \\
		\hline
		\bf M2e4\_R10\_mu0.42 & 2 & 0.42 & 16 & 
		$2\times 10^4$ & 16 & 10 & 200 & 
		0.008 & 4 & 1 &  0.0078 & 2 & 
		$3\times 10^{-3}$ & $7 \times 10^{-5}$ & 1.4 &  $2\times10^{7}$ \\
		\hline
		\end{tabular}
        \vspace{-0.1cm}
 \caption{Initial conditions of clouds used in our runs (see \S~\ref{sec:params_scales} for definitions). The scaled parameters give the properties of a corresponding physical GMC model with $\Sigma_\mathrm{gas} \sim 63 M_\odot\,\rm pc^{-2}$, typical in the Milky Way, with $M_0$ being the initial cloud mass. Note that the parameters in theis table apply to both \textit{Box} and \textit{Sphere} runs as they set up to have identical initial global parameters, with  $L_{\mathrm{box}}$ being the box size for \textit{Box} runs and $R_{\mathrm{cloud}}$ the being the cloud radius for the \textit{Sphere} runs. Note that \textit{Box} runs have slightly different initial parameters (e.g., Mach number, virial parameter) due to the non-exact scaling of the driving, so the values shown here are the target values. Also, different works in the literature use different Jeans mass definitions, which can change $\MJeans$ up to a factor of 10, ours is defined by Eq. \ref{eq:mjeans_simple}.}
 \label{tab:IC}\vspace{-0.5cm}
\end{table*}

 
 \section{Results}\label{sec:results}

\begin{figure*}
\begin {center}
\includegraphics[width=0.99\linewidth]{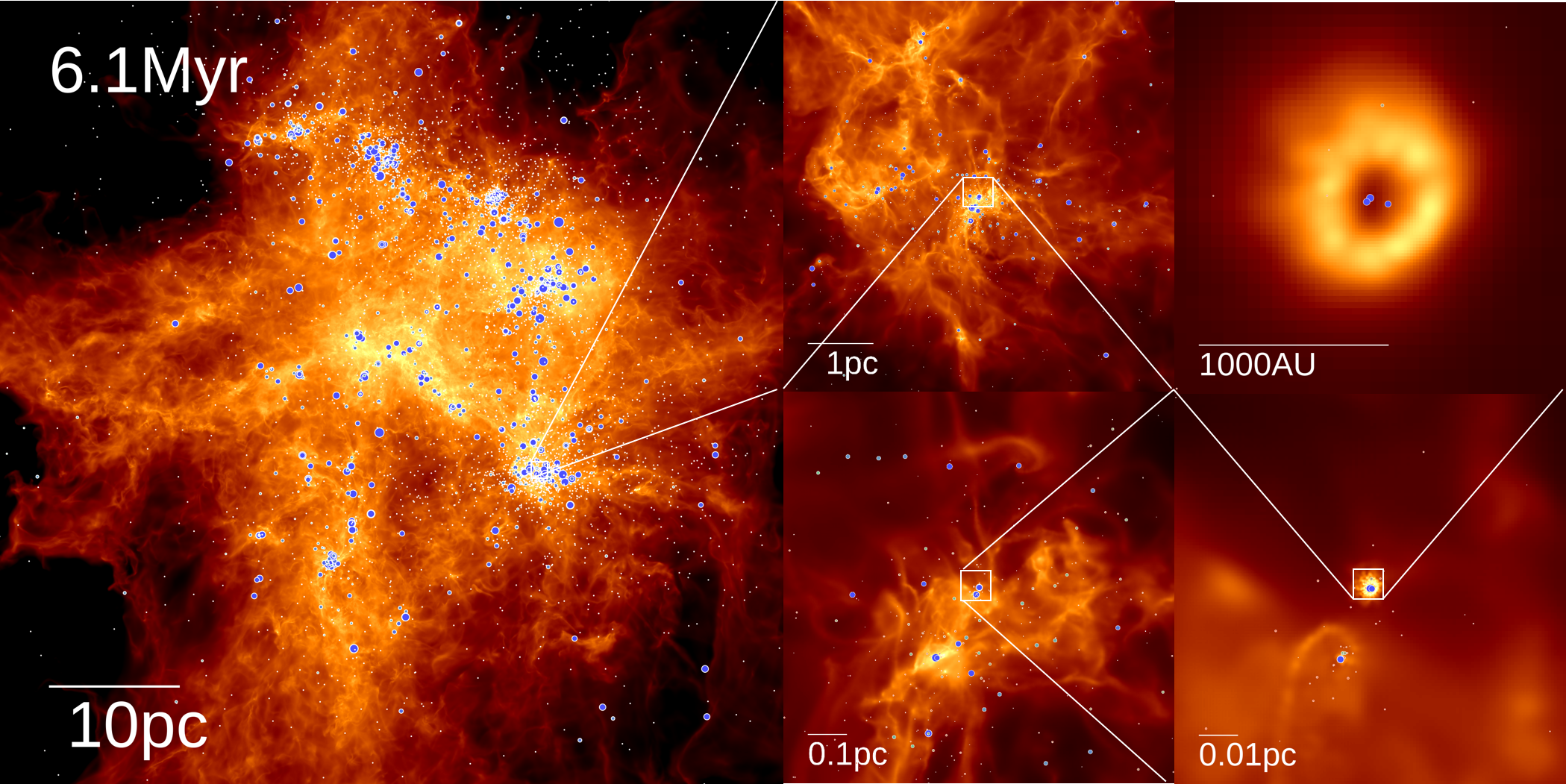}
\caption{Surface density maps from a simulation of a $2\times 10^5\,\msun$ GMC that includes isothermal turbulence and MHD (\textbf{M2e5\_R30}, see Table \ref{tab:IC}), at about 8\% star formation efficiency. The color scale is logarithmic and the circles represent sink particles (stars) that form in high-density regions where fragmentation can no longer be resolved, their size increasing with mass. This simulation resolves a dynamic range from $\sim\!\mathrm{50\,pc}$ down $\sim\!\mathrm{30\,AU}$.}
\label{fig:starforge}
\vspace{-0.5cm}
\end {center}
\end{figure*}

 We carried out a suite of simulations in the $\alphaturbzero$-$\mach$-$\mu$ parameter space at various resolutions, up to $M_0/\Delta m=2\times 10^{8}$ (see Table \ref{tab:IC} for details and Figure \ref{fig:starforge} for a demonstration of the dynamic range). This is the highest mass resolution yet achieved in any 3D simulation of resolved star cluster formation.

\begin{figure*}
\begin {center}
\includegraphics[width=0.99\linewidth]{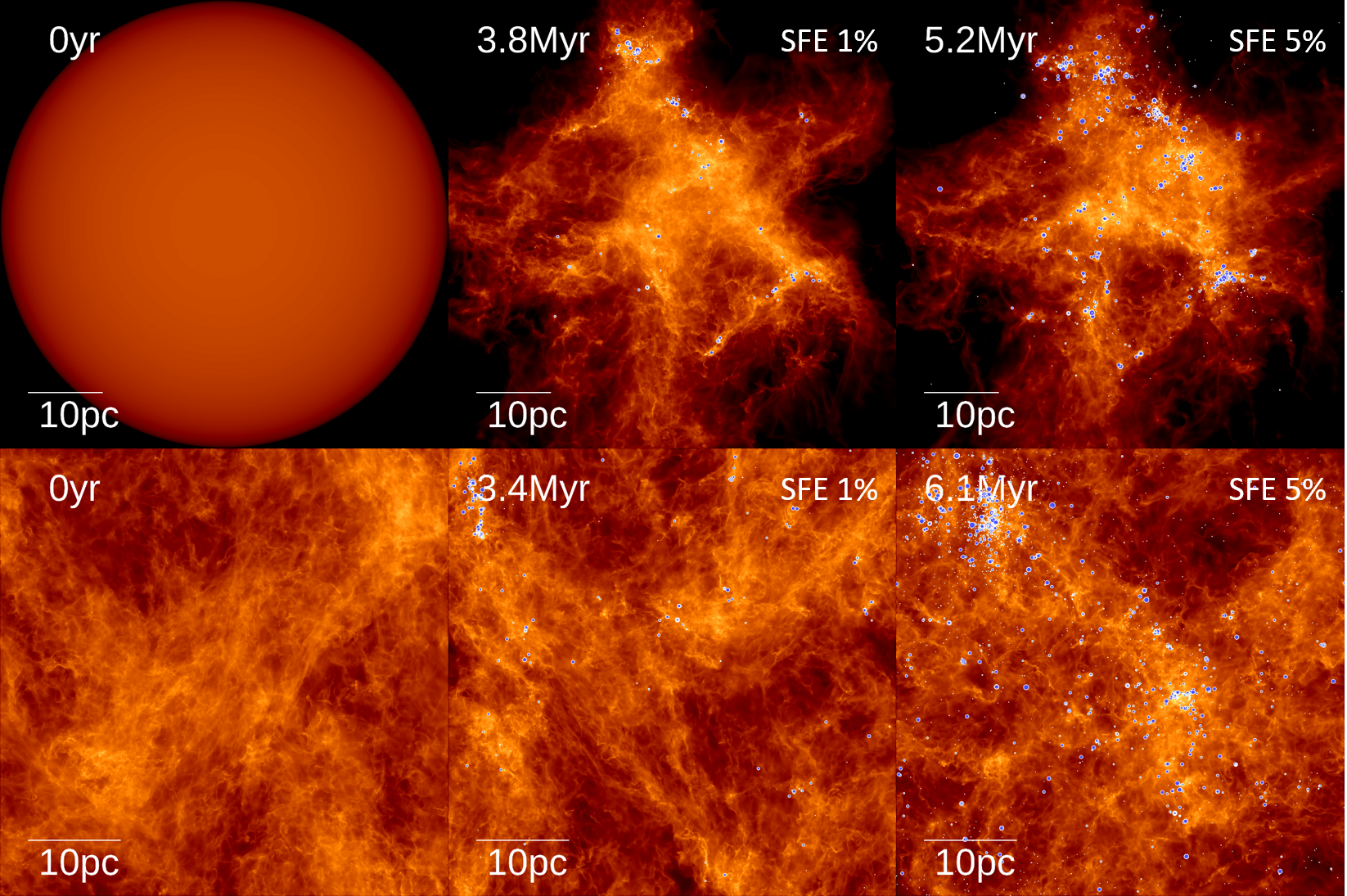}
\caption{Surface density maps for the same cloud type as Figure \ref{fig:starforge} (\textbf{M2e5\_R30}, see Table \ref{tab:IC}), both for \textit{Sphere} and \textit{Box} initial conditions (top and bottom row respectively), when the simulation starts and at 1\% and 5\% star formation efficiency (columns, left to right).}
\label{fig:starforge_series}
\vspace{-0.5cm}
\end {center}
\end{figure*}

\begin{figure}
\begin {center}
\includegraphics[width=0.99\linewidth]{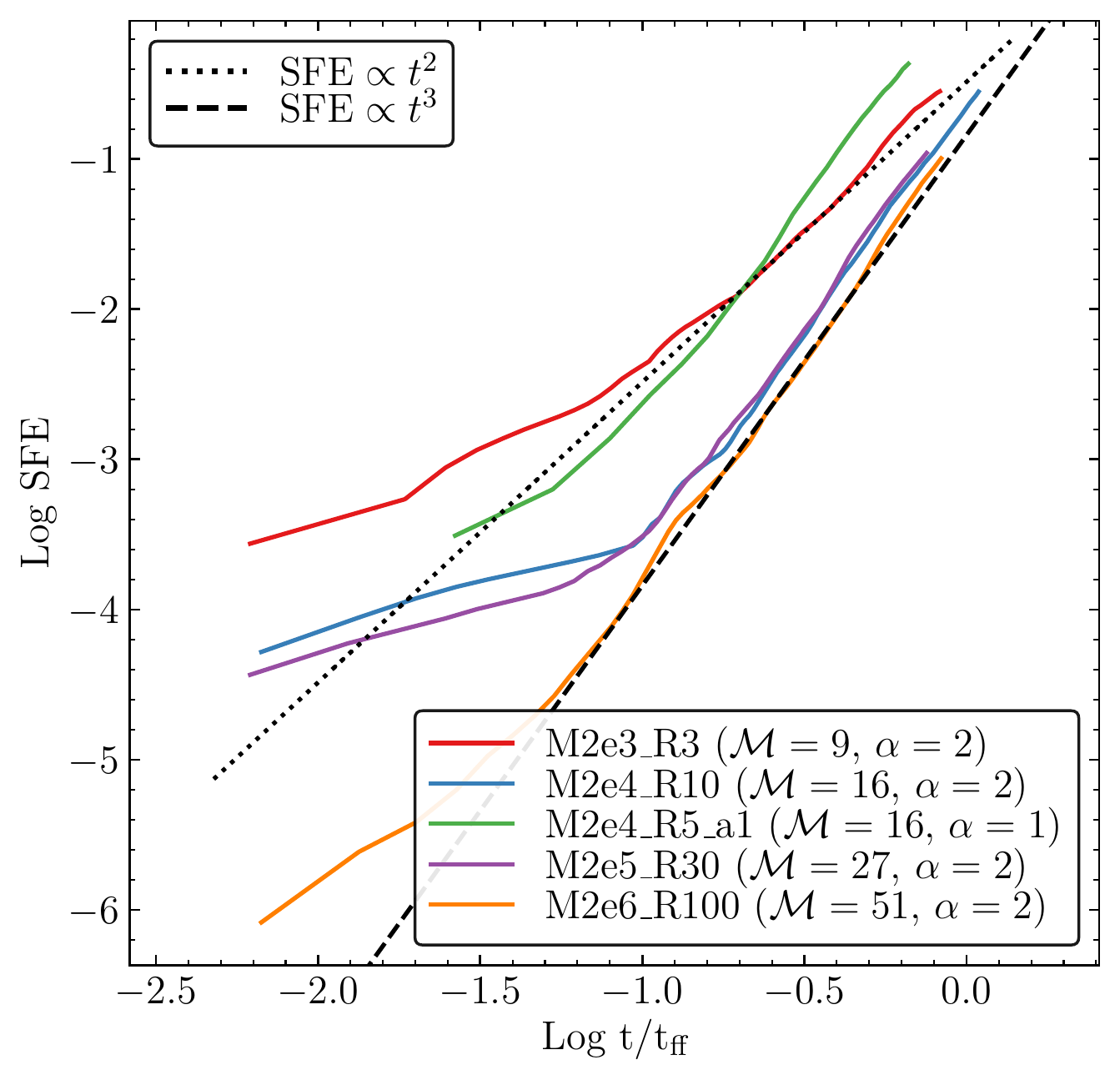}
\caption{Evolution of the star formation efficiency ($\mathrm{SFE}(t)=M_{\mathrm{sink}}(t)/M_{0}$) as function of time for a subset of runs. SFE rises as a broken power-law of time and reaches about 10\% in about one freefall time $\left(t_{\mathrm{ff}}=\sqrt{\frac{3 \pi}{32 G \rho_0}}\right)$.}
\label{fig:sfe_t_evol}
\vspace{-0.5cm}
\end {center}
\end{figure}

Once the simulation begins we find that the clouds quickly develop a filamentary structure similar to observations (\protect\citealt{Andre_2010_filaments}) that collapses and forms stars (see Figure \ref{fig:starforge_series}). Figure \ref{fig:sfe_t_evol} shows that all our clouds turn roughly 10\% of their gas into stars in a freefall time. At low Mach numbers ($\mach<10$) we find a rough trend of $\mathrm{SFE}\propto t^2$ (consistent with the results of \citealt{lee:2015.mhd.sf} who simulated a $\mach=9$ cloud), while for all highly supersonic clouds ($\mach>10$) the relation becomes steeper, consistent with $\mathrm{SFE}\propto t^3$. This does not necessarily contradict the theory of \citet{murray_star_formation}, who derived $\dot{M}_\star \propto t^2$ for a {\it single} star accreting in a turbulent medium -- our star formation history is the sum of many individual stellar accretion histories.

\subsection{Sink mass distribution (IMF)}

\begin{figure*}
\begin {center}
\includegraphics[width=0.49\linewidth]{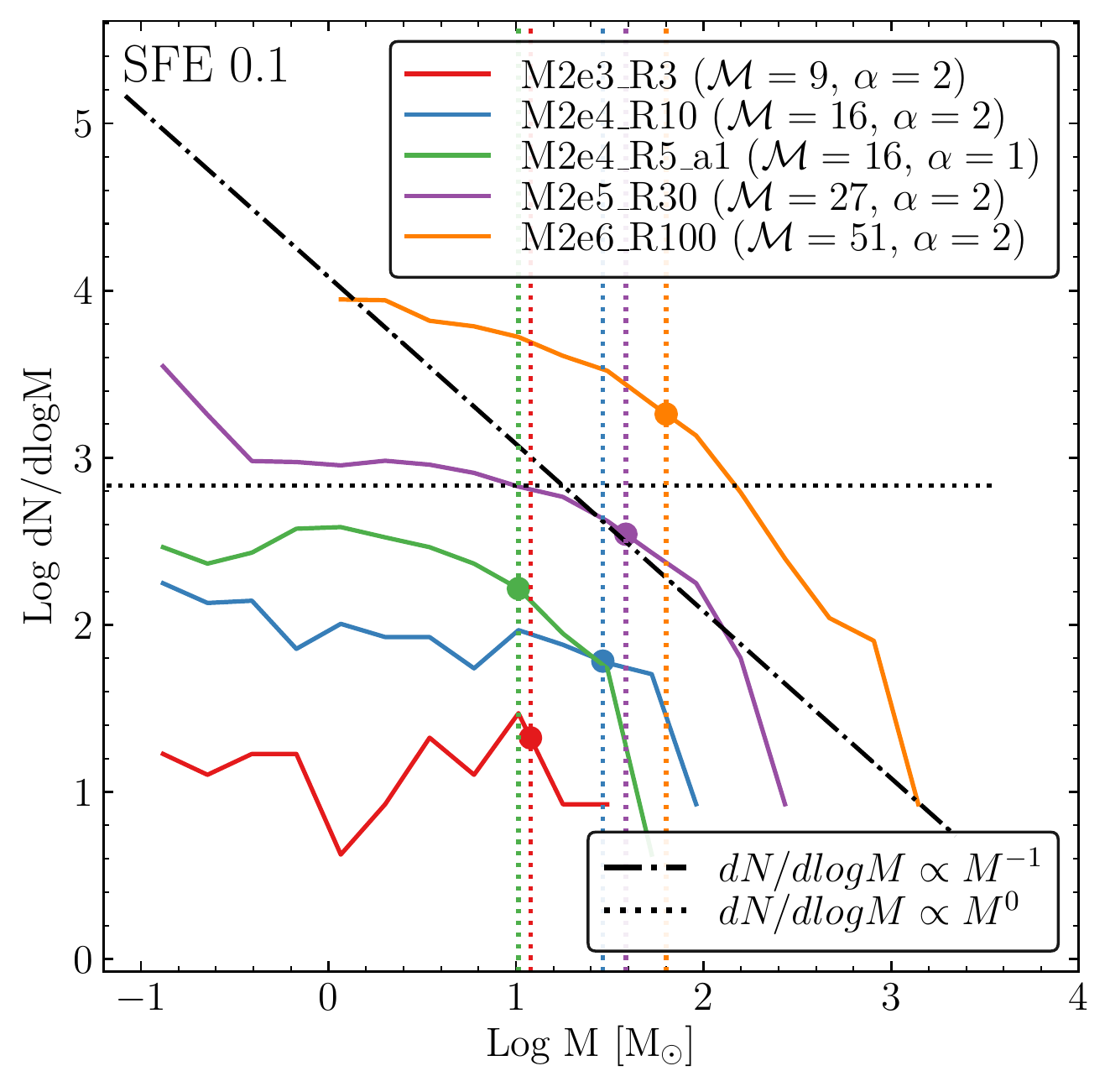}
\includegraphics[width=0.49\linewidth]{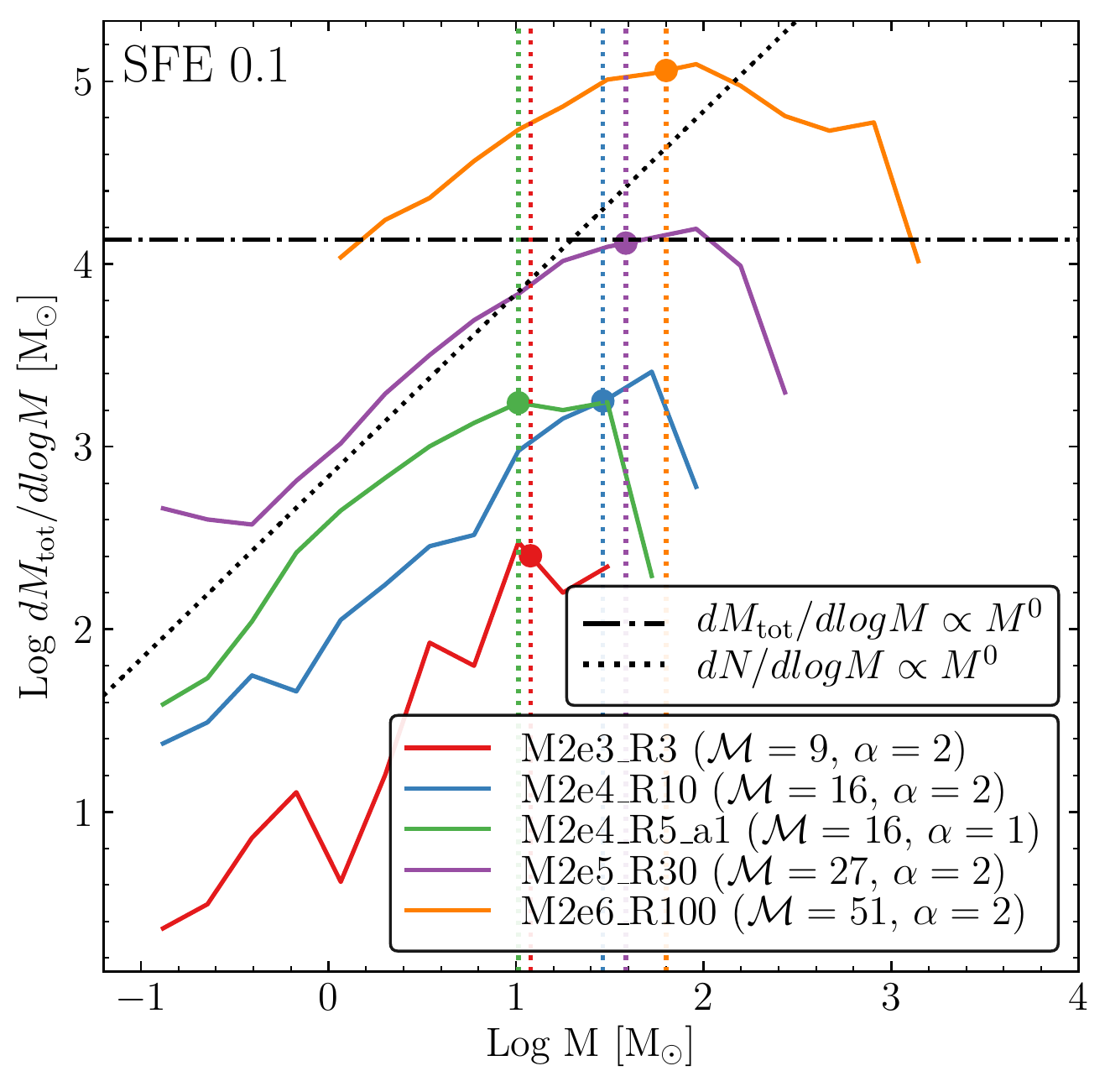}
\vspace{-0.4cm}
\caption{Distribution of sink particle masses at 10\% star formation efficiency ($\mathrm{SFE}=\sum (M_{\mathrm{sink})}/M_{0}$) for a subset of our runs using \textit{Sphere} initial conditions. The chosen runs have $\alphaturbzero=2$ and a similar mass-size relation to observed MW GMCs, except for one that has $\alphaturbzero=1$ (see Table \ref{tab:IC}). The dotted vertical lines and the circular symbols denote the mass-weighted median sink mass $\Mmedian$, while the dash-dotted and dotted lines are the analytical results for a Salpeter-like $\dderiv N/\dderiv \log M\propto M^{-1}$ and a shallower $\dderiv N/\dderiv \log M=\mathrm{const.}$ sink population distributions. To make the plots easier to parse the \textit{y} axes are not normalized. Note that we only plot sink particles more massive than 100 times the mass resolution, as results below that might be sensitive to our choice of sink particle algorithm. \textit{Left:} The mass PDF of sink particles ($\dderiv N_{\mathrm{sink}}/\dderiv \log M$) . The PDF rises steeply at the high mass end then turns over to a flat distribution. \textit{Right:} Distribution of mass among sink particles for the same runs ($\dderiv M_{\mathrm{sink}}/\dderiv \log M \sim M\,\dderiv N/\dderiv\log M$). At high masses the distribution is flat (consistent with $\dderiv N/\dderiv \log M\propto M^{-1}$) then becomes linear ($\dderiv N/\dderiv \log M=\mathrm{const.}$). Note that $\Mmedian$ roughly corresponds to the point where the slope of the power law changes, similar to the turnover mass in the observed IMF. For the rough scaling of $\Mmedian$ with $\mach$ and $\alphaturbzero$ see Equation \ref{eq:m50_scaling_no_mass}.}
\label{fig:stat_examples}
\vspace{-0.5cm}
\end {center}
\end{figure*}

Figure \ref{fig:stat_examples} shows that varying the initial conditions (in this case the virial parameter $\alphaturbzero$ and Mach number $\mach$) significantly changes the mass distribution of sink particles. At high masses the sink distribution is consistent with a $\dderiv N/\dderiv \log M\propto M^{-1}$ power law, similar to the observed IMF \citep{salpeter_slope, imf_universality}. Meanwhile, at low masses the distribution becomes shallower, consistent with $\dderiv N/\dderiv \log M\sim\mathrm{const}$. This is significantly shallower than the low mass end of the observed IMF ($\dderiv N/\dderiv \log M\sim M^{0.7}$ in the \citealt{kroupa_imf} form), leading to an excess of brown dwarfs, which should only make up $\sim 30\%$ of the stellar population \citep{andersen_2006_browndwarfs}. Meanwhile, the turnover from the high mass power-law behavior shows that the sink mass distribution does have a mass scale inherited from initial conditions. For simplicity we adopt the mass-weighted median mass of sinks $\Mmedian$ as the characteristic mass scale of sinks in our subsequent analysis (similar to \citealt{krumholz_2012_orion_sims}), as it roughly corresponds to this turnover mass (see Figure \ref{fig:stat_examples}). This characteristic mass $\Mmedian$ monotonically increases as more gas is turned into stars (see Figures \ref{fig:box_vs_sphere} and \ref{fig:M50_sfe_evol} for values).
 
\subsection{Effects of turbulent driving and boundary conditions (\textit{Box} vs \textit{Sphere})}\label{sec:box_vs_sphere}

\begin{figure}
\begin {center}
\includegraphics[width=\linewidth]{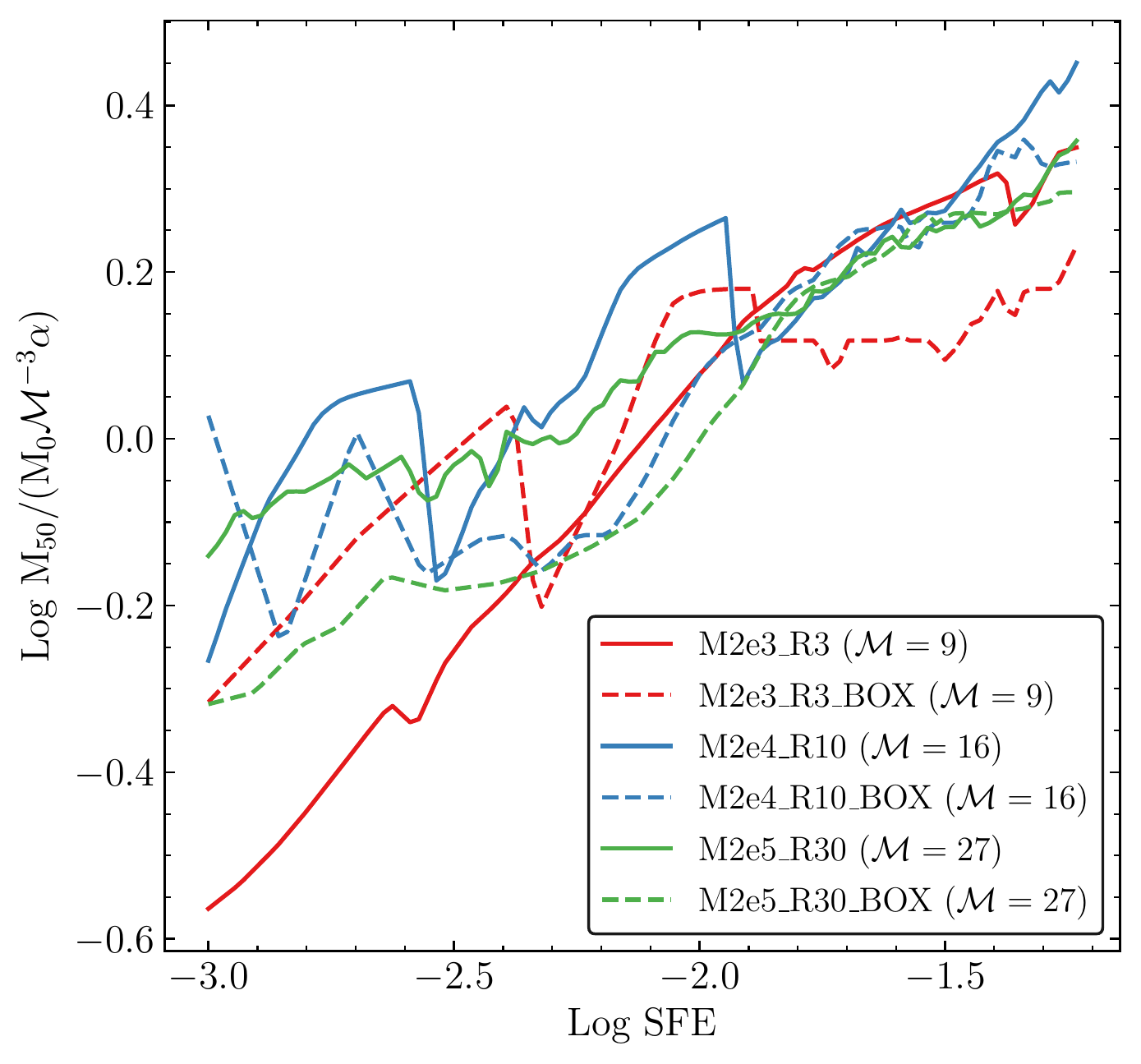}
\vspace{-0.4cm}
\caption{The mass-weighted median sink mass (normalized to our approximate best-fit scaling as a function of cloud mass, initial Mach number and turbulent virial parameter) as a function of star formation efficiency (see Eq. \ref{eq:m50_scaling} for details on the scalings). We find no clear difference between runs with \textit{Box} and \textit{Sphere} initial conditions.}
\label{fig:box_vs_sphere}
\vspace{-0.5cm}
\end {center}
\end{figure}

While the global parameters of the initial conditions ($\alphaturbzero$, $\mach$, $M_0$) affect the mass spectrum of sink particles, we find no significant difference between \textit{Sphere} and \textit{Box} runs (see Figure \ref{fig:box_vs_sphere}), despite the difference in initial cloud shape, turbulent driving, density and magnetic fields\footnote{It should be noted that while the exact magnitude of magnetic support on large scales appears to be irrelevant, having finite (non-zero) magnetic fields is crucial because, in the limit of no magnetic fields, clouds undergo an infinite fragmentation cascade, see \S~\ref{sec:mu_insensitivity} and \citet{guszejnov_isothermal_collapse} for details.}\footnote{Note that we use $\alphaturbzero$ based on Eq. \ref{eq:alphaturb_sphere} similar to other studies in the literature. For a periodic box this is can significantly differ from the value $\alphaturb$ from Eq. \ref{eq:alpha_turb} \citep{federrath_sim_2012}.} The insensitivity of the sink mass spectrum to the specifics of the initial conditions is similar to the findings of \protect\citet{Bate_2009_IC_test}, \protect\citet{Liptai_2017_IMF_from_turb_IC} and \protect\citet{Lee_Hennebelle_2018_IC}. 

Note that studies simulating dense, centrally concentrated clouds found that the final sink masses depend on the initial condition \citep{Girichidis_2011_isoT_sim}. These initial conditions, however, are quite different from what is observed in GMCs. Furthermore, \cite{Girichidis_2011_isoT_sim} simulated isothermal turbulence without magnetic fields, which have been shown to produce sink mass spectra entirely set by numerical resolution \citep{guszejnov_isothermal_collapse}.

Previous studies have shown that the driving mode of turbulence has significant effect on the star formation histories of clouds (e.g., \citealt{federrath_sim_compare_2010}), which is apparent in our results as well (see Figure \ref{fig:starforge_series} for an illustration). But since we found the mass-weighted median sink mass $\Mmedian$ to be insensitive to even whether there is driving or not, we left the exploration of the effects of different driving modes to a future study.

\subsection{$\Mmedian$ as a function of initial conditions}

Neglecting variations with $\mu$, we find that the evolution and parameter-dependence of $\Mmedian$ is well-described by the following formula:

\be
\Mmedian/M_0 =  7.8 \left(\mathrm{SFE}\right)^{0.3} \mach^{-3.2} \alphaturbzero^{1.1} \pm 0.06\mathrm{dex}, 
\label{eq:fitting_func}
\ee
where the parameters and the overall RMS fitting error were obtained from an unweighted least-squares fit to all simulations with our fiducial $\mu=4.2$, excluding snapshots with $<5$ sink particles and with $\mathrm{SFE}>0.1$. 
This fit appears to collapse all simulations to a single curve, with no obvious trend in the residuals with any of the dimensionless parameters (see Appendix \ref{sec:fitting} for details). The runs that deviate most from the best-fit relation happen to be the lower-$\mathcal{M}$ clouds that produce the smallest number of sinks at fixed SFE, suggesting that the deviations are simply statistical noise from the \myquote{sampling} process of the underlying IMF.


Based on this fit, the rough scaling of the characteristic mass $\Mmedian$ (at fixed SFE) is
\be 
\Mmedian \appropto M_0 \mach^{-3} \alphaturbzero.
\label{eq:m50_scaling}
\ee
This is similar to the scalings of both $\Msonic$ and $\MJeans$ (see Eqs. \ref{eq:mjeans_simple}-\ref{eq:msonic_simple}), but neither of those matches our results exactly (see Appendix \ref{sec:fitting}). Assuming the existence of a mass-size and a linewidth-size relation similar to that in the MW ($M_0\propto R_0^{2}$ and $\mach^2 \propto R_0$ respectively, see \citealt{larson_law}), we can eliminate the cloud mass $M_0$ and rewrite Equation \ref{eq:fitting_func} as
\be 
\Mmedian \appropto \mach \alphaturbzero^{-1},
\label{eq:m50_scaling_no_mass}
\ee
 see Figure \ref{fig:stat_examples} for an illustration of the scaling with $\mach$.
 
 In dimensional units, in terms of the cloud mass $M_\mathrm{0}$, surface density $\Sigma=M_\mathrm{0} / \pi R_\mathrm{cloud}^2$, SFE, and virial parameter $\alpha_\mathrm{turb}$, our fit of Equation \ref{eq:fitting_func} can be expressed as
\begin{equation}
\Mmedian \approx 24 M_\odot\, \left(\frac{\mathrm{SFE}}{0.05}\right)^{0.3}\,M_\mathrm{5}^{0.2} \,\alphaturbzero^{-0.5} \, \Sigma_{100}^{-0.8} \, c_\mathrm{s,0.2}^{3.2},
\label{eq:m50_scaling_MW}
\end{equation}
where $M_\mathrm{5}= \frac{M_\mathrm{0}}{10^5M_\odot}$, $\Sigma_\mathrm{100}= \frac{\Sigma}{100\,M_\odot\,\mathrm{pc}^{-2}}$, and $c_\mathrm{s,0.2}=\frac{c_\mathrm{s}}{0.2\mathrm{km\,s^{-1}}}$, normalizing to typical values for GMCs in the Milky Way \citep[e.g.][]{larson_law}.

Using the same procedure as with $\Mmedian$ in Eq. \ref{eq:fitting_func}, we also fit the maximum stellar mass $M_\mathrm{\star max}$, obtaining
\begin{equation}
    M_\mathrm{\star,max}/M_\mathrm{0} = 1.2 \left(\mathrm{SFE}\right)^{0.5} \mach^{-1.8} \alphaturbzero^{0.5} \pm 0.1\mathrm{dex}.
    \label{eq:mmax_fit}
\end{equation}


\subsection{(In-)sensitivity of $\Mmedian$ to $\mu$}\label{sec:mu_insensitivity} 
\begin{figure}
\begin {center}
\includegraphics[width=0.99\linewidth]{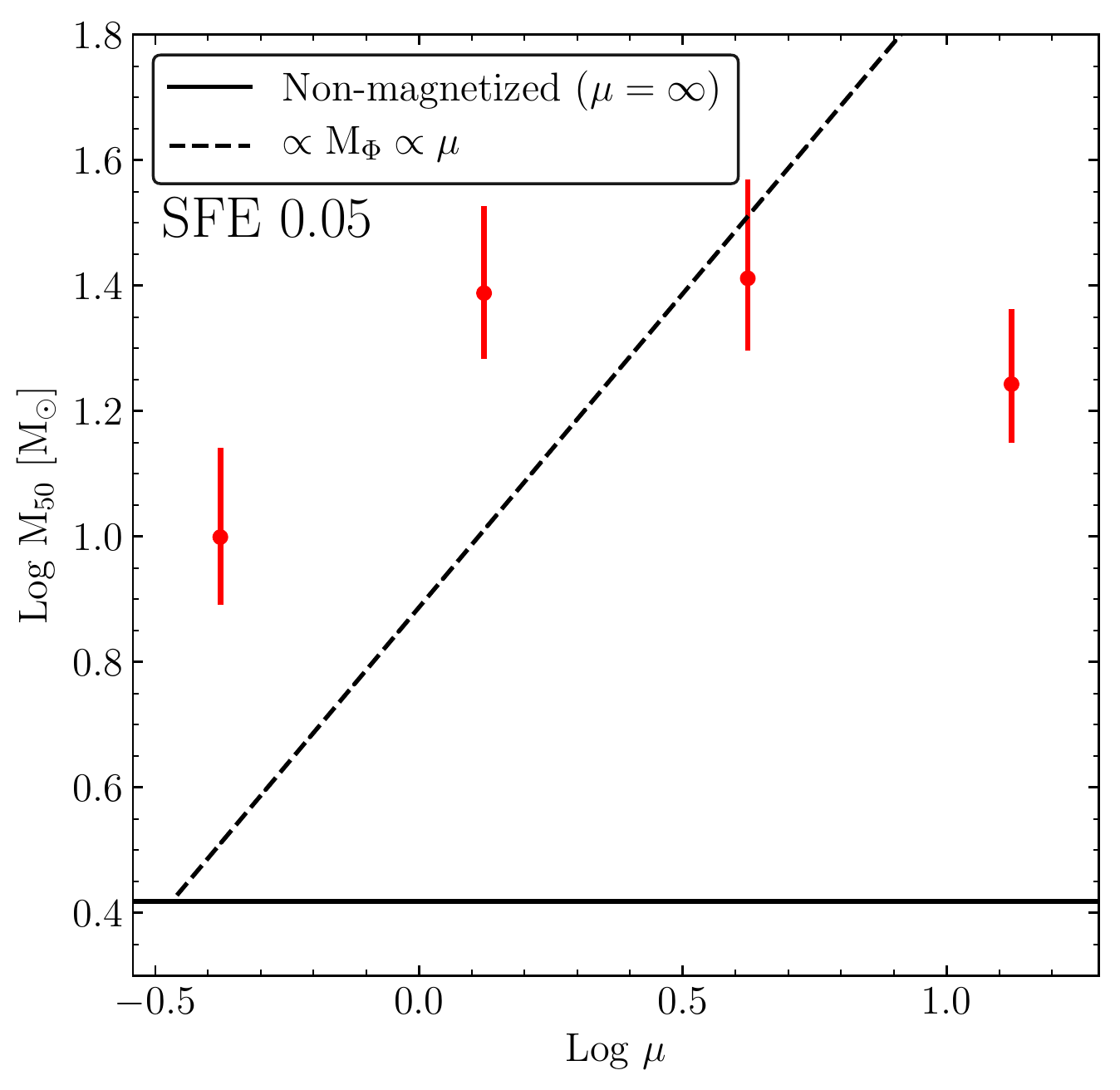}
\vspace{-0.4cm}
\caption{Dependence of the mass-weighted median sink mass $\Mmedian$ at 5\% SFE on the normalized mass-to-flux ratio $\mu$ (or equivalently, mean magnetic field strength) in a $2\times10^4\,\msun$ GMC (e.g., \textbf{M2e4\_R10\_mu1.3}, see Table \ref{tab:IC}). Recall, for otherwise equal parameters, $\mu \propto 1/B_{0}$ is inversely proportional to the mean magnetic field strength. The solid black line shows $\Mmedian$ in a non-magnetized run (at the same resolution), while the dashed line shows the expected behavior if $\Mmedian$ was set by the magnetic critical mass (Equation \ref{eq:M_Phi}). We only include Sphere runs as the magnetic field energy saturates a constant fraction of the kinetic energy in the \textit{Box} ICs \citep{federrath_2014_dynamo}. The errors are estimated by bootstrapping: we resample the sink mass distribution at fixed total stellar mass and calculate the 95\% confidence interval of the mass-weighted median mass over these new realizations.} 
\label{fig:M50_dependence_mu}
\vspace{-0.5cm}
\end {center}
\end{figure}

An interesting aspect of our results is that $\Mmedian$ appears to be insensitive to the initial magnetic field strength (see Figure \ref{fig:M50_dependence_mu}), but without magnetic fields we have found that clouds fragment without limit, making $\Mmedian$ dependent on numerical resolution \citep{guszejnov_feedback_necessity, guszejnov_isothermal_collapse}.

\begin{figure}
\begin {center}
\includegraphics[width=0.99\linewidth]{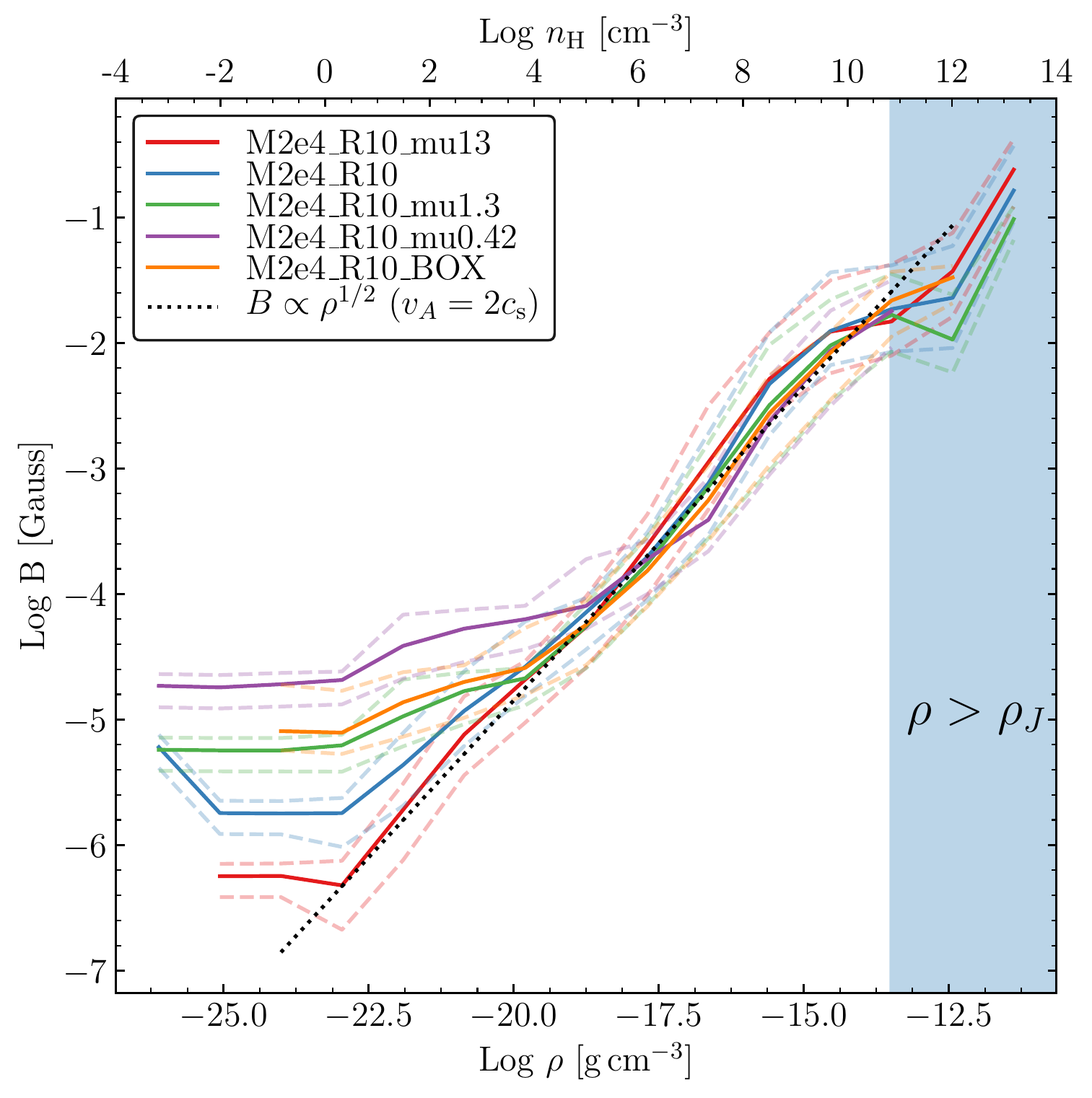}
\vspace{-0.4cm}
\caption{Magnetic field strength as a function of gas density in the \textbf{M2e4\_R10} runs at the same $\Delta m=0.001\,\msun$ mass resolution with different initial magnetic fields and ICs (see Table \ref{tab:IC}) at 5\% SFE. The solid lines show the mass weighted median of the magnetic field in different density bins (equivalent to median value for cells as MFM cells have equal masses), while the dashed lines show the 25th and 75th percentiles. The shaded region marks densities exceeding the maximum Jeans-resolved density $\rho_J$ (Equation \ref{eq:rhoJ}). To achieve satisfactory statistics at the high density end we stacked the distribution from 10 snapshots around the target SFE. Despite the different initial conditions all runs saturate to the same $B\propto \rho^{1/2}$ line (corresponding to $v_{A}=2\cs$), similar to the results of \citet{Wurster_2019_no_magnetic_break_catastrophe}. The results depart from  the power-law above $\rho \sim 3\times 10^{-14}\mathrm{g\,cm}^{-3}$, corresponding to the maximum Jeans-resolved density $\rho_J$ for these simulations (Equation \ref{eq:rhoJ}).} 
\label{fig:B_rho}
\vspace{-0.5cm}
\end {center}
\end{figure}

Figure \ref{fig:B_rho} shows that regardless of the initial magnetic field strength, the turbulent dynamo in the system drives the systems towards a {\em common} $B-\rho$ relation at high densities. This is in good agreement with the findings of \citet{mocz_2017_core_sim, Wurster_2019_no_magnetic_break_catastrophe, Lee_Hennebelle_2019_T_B}, who, using different numerical schemes, find the $B$-$\rho$ relation to saturate to the same trend, regardless of initial magnetic field strength. Furthermore, we find that this result is insensitive to not only the initial field strength but also to whether we have decaying (\textit{Sphere}) or driven (\textit{Box}) turbulence in the simulation. 

It is unclear what exactly causes the $B\propto\rho^{1/2}$ relation observed in our simulations (see Figure \ref{fig:B_rho}). A possible explanation of the exponent is that it arises from the anisotropic collapse of magnetic flux-conserving gas in both disk-like and cylindrical geometries \citep[see][]{Tritsis_2015_B_rho_relation}. One problem with this interpretation is that both our results and the ones in the literature saturate to the \emph{same} relation, regardless of the initial field strength (as opposed to parallel ``tracks,'' which is what one would obtain for different initial $\mu$ values in a pure flux-freezing argument). What is striking is that this universal normalization roughly corresponds to $v_A(\rho)\sim 2\cs$, where $v_A(\rho)$ is the local Alfvén velocity at density $\rho$. This is suspiciously close to equipartition. One possibility is that the normalization of the $B$-$\rho$ relation is enforced by a local dynamo effect (similar to the global $\alphaB$ saturating in driven boxes, see \citealt{federrath_2011_dynamo}) that is driven by the local gravitational collapse. In numerical experiments, $\beta \sim 1$ is generally achieved for trans- or modestly super-sonic turbulence \citep{stone_1998_mhd_dynamo}, which was indeed found on all scales in individual collapsed cores by \citet{mocz_2017_core_sim}.

Of course, if the initial magnetic field was much larger than the ``saturation'' values predicted here at high densities, this would alter out conclusions, but such large fields would imply the initial cloud is not self-gravitating at all.

A local small-scale dynamo effect would also explain why our isothermal MHD results, although insensitive to the exact initial value of the magnetic field strength, are qualitatively different from our previous isothermal non-MHD results \citep{guszejnov_isothermal_collapse}. If magnetic fields are present, they are amplified to this line, regardless of their initial value, and prevent the fragmentation cascade that would happen in the non-magnetized case.

In Figure \ref{fig:B_rho} we also note a departure from the $B \propto \rho^{1/2}$ relation above $\rho \sim 3\times 10^{-14} \mathrm{g\,cm}^{-3}$, which corresponds to the maximum density at which the smallest unstable Jeans modes can possibly be resolved, $\rho_J$ (Equation \ref{eq:rhoJ}). We have verified that this departure from power-law behaviour is an artifact of the finite resolution of the simulations ($\Delta m=10^{-3} \msun$), as our version of {\bf M2e4\_R10} at our maximum resolution of $10^{-4}\msun$ has a similar turn-over at $\sim 100\times $ higher density. This deficit of magnetic energy at densities $>\rho_J$ may be due to a numerical suppression of small-scale energy injection through gravitational collapse at the smallest unstable Jeans scale, which would otherwise drive turbulence and the small-scale dynamo in turn \citep{federrath_2011_jeans_criterion}. 

\subsection{Resolution insensitivity of the characteristic mass}

\begin{figure}
\begin {center}
\includegraphics[width=0.99\linewidth]{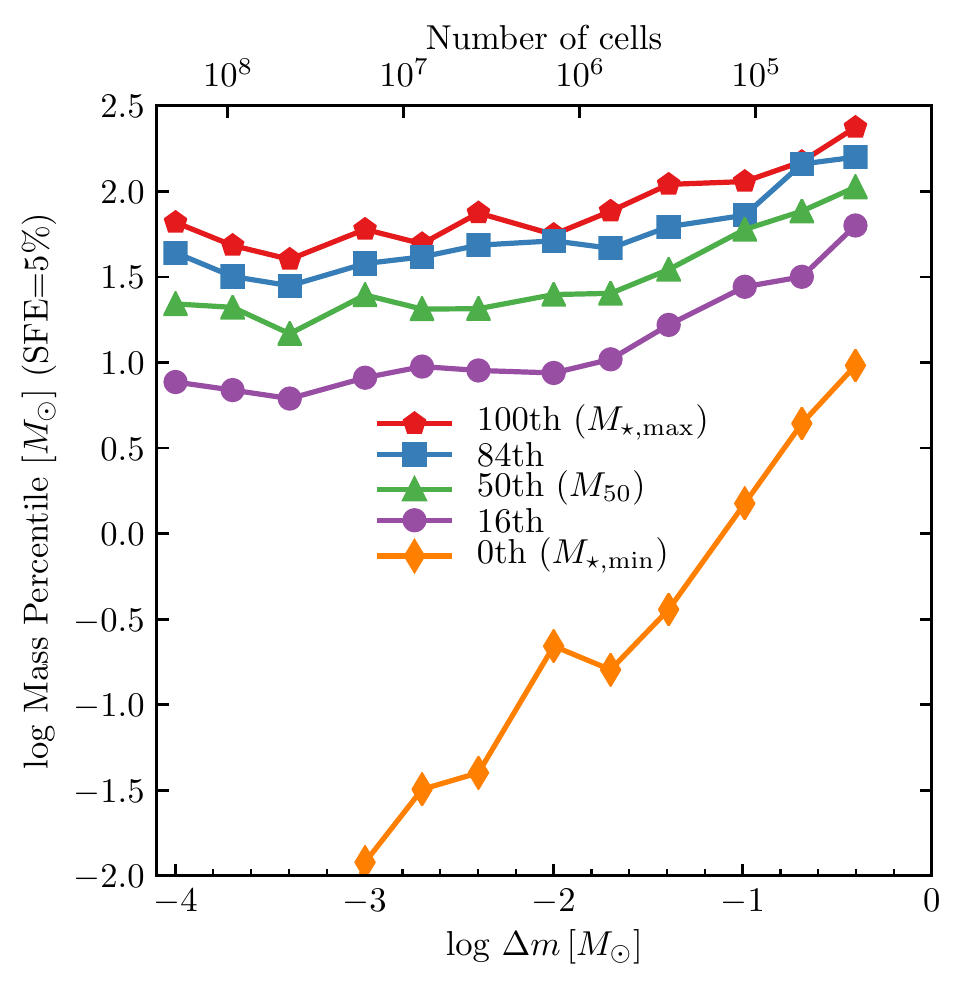}
\vspace{-0.4cm}
\caption{``Percentile curves'' showing the sink mass below which $X\%$ of the total mass in sinks resides (measured when each simulation has reached a SFE of $5\%$), in otherwise-identical {\bf M2e4\_R10} simulations as a function of mass resolution $\Delta m$, corresponding to the number of Lagrangian gas cells in the cloud (top axis). The ``50th'' curve is just the mass-weighted median $M_{50}$ as defined above, while 0th and 100th are the minimum and maximum sink mass in the simulation. The minimum mass scales proportional to $\Delta m$ because the predicted IMF has no discernible lower cutoff (Figure \ref{fig:stat_examples}). However, the higher percentiles appear to become insensitive to resolution for sufficiently low $\Delta m$ (high resolution).}
\label{fig:percentiles}
\vspace{-0.5cm}
\end {center}
\end{figure}

In the non-magnetized case, clouds fragment to infinitely small scales as discussed in \S~\ref{sec:intro} and in \citet{guszejnov_isothermal_collapse}, so any apparent mass scale in the sink mass distribution is inescapably tied to numerical resolution. It is therefore crucial to check for the resolution dependence of $\Mmedian$. Figure \ref{fig:percentiles} shows how various mass-weighted percentiles of the IMF vary as a function of mass resolution for the {\bf M2e5\_R30} run. The {\it minimum} stellar mass continuously decreases $\propto \Delta m$, but the {\it maximum} stellar mass, and the intermediate mass-weighted percentiles (i.e. stellar mass below which there is X\% of the total {\it mass} in the IMF), level off above a certain resolution threshold. 

\begin{figure}
\begin {center}
\includegraphics[width=0.99\linewidth]{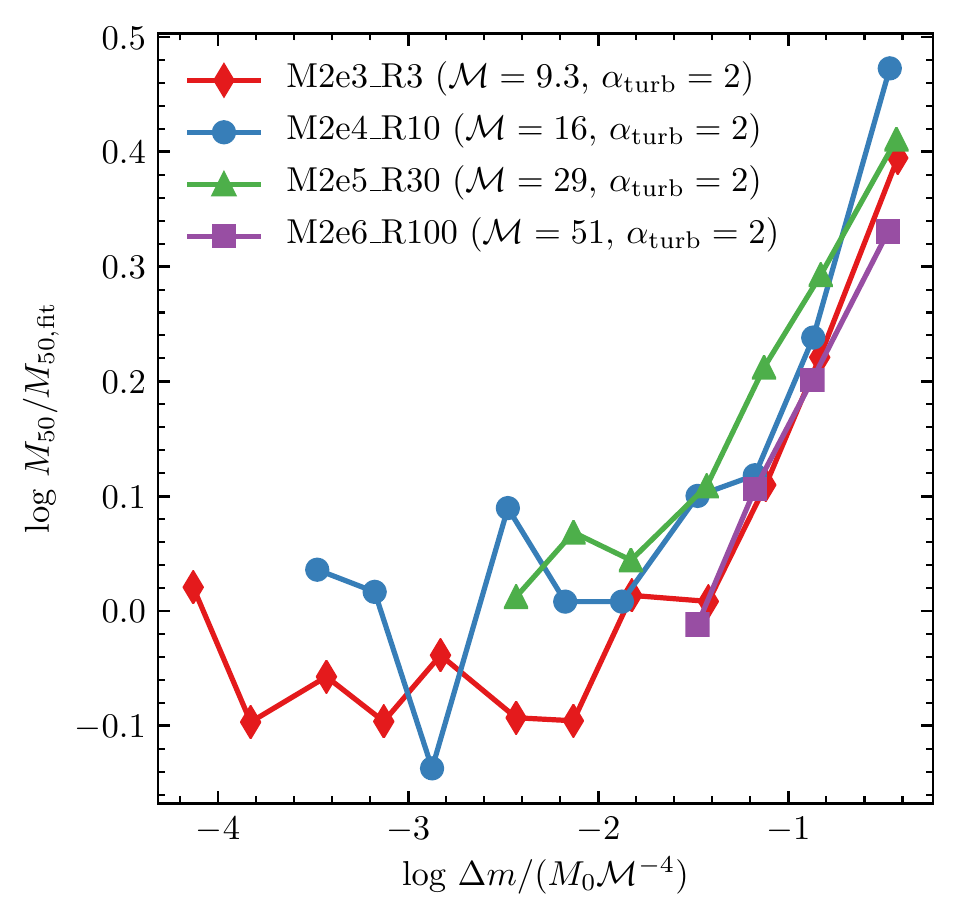}
\vspace{-0.4cm}
\caption{Mass resolution dependence of the predicted mass-weighted median stellar mass $M_{50}$ in {\it Sphere} runs at 5\% SFE, rescaled to the best-fit value to the respective highest available resolution levels (Eq \ref{eq:fitting_func}). When the resolution criterion $\Delta m << M_\mathrm{0} \mach^{-4}$ is satisfied (Equation \ref{eq:convergence_crit}), the predicted $M_{50}$ becomes insensitive to $\Delta m$.} 
\label{fig:convergence}
\vspace{-0.5cm}
\end {center}
\end{figure}

For this specific model ({\bf M2e4\_R10}), the apparent resolution criterion is $\Delta m \approx 3 \times 10^{-8} M_0 = 0.01\,\msun$, however the problem is scale-free, so we expect that the resolution criterion will more generally assume the form $\Delta m \leq M_0 \mach^{p_1} \alphaturbzero^{p_2} \mu^{p_3}$, for some exponents $p_1$, $p_2$, and $p_3$. Lacking a detailed convergence study for runs that vary $\alphaturbzero$ and $\mu$, we focus on the criterion for simulations with the fiducial values of these parameters (2 and 4.2, respectively). For all runs, at all times and SFE, we have examined the variation of $\Mmedian/M_{\mathrm{50,\infty}}$ as a function of mass resolution, where $M_{\mathrm{50,\infty}}$ is the value obtained in the limit $\Delta m \rightarrow 0$. In practice we use the value given by Equation \ref{eq:fitting_func} as a proxy for $M_{\mathrm{50,\infty}}$, which is a fit to the respective highest available resolution levels for each simulation. Figure \ref{fig:convergence} shows that with increasing resolution $M_{50}$ approaches the value given by Equation \ref{eq:fitting_func}. This value is reached in {\em all} simulations when the following criterion is satisfied:
\be
\Delta m \lesssim 0.05 M_\mathrm{0} \mach^{-4}.
\label{eq:convergence_crit}
\ee


For $\alphaturbzero \sim 1$, this is simply the criterion that the sonic mass (Equation \ref{eq:msonic_simple}) or turbulent Bonnor-Ebert mass (Equation \ref{eq:MBE_simple}) be resolved by $\gtrsim 20 \Delta m$. These are both proposed characteristic core masses in turbulent fragmentation \citep{padoan_2007, core_imf}, and the specific number is on the order of the minimum number of Lagrangian mass elements for the stability of a clump to be insensitive to numerical discretization and softening details (\citealt{Bate_1995_accretion, price_monaghan_softening}, Grudi\'{c} et al. 2020, in prep.). Thus Equation \ref{eq:convergence_crit} simply expresses the requirement that {\em the collapse of gravitationally-unstable cores formed via turbulent fragmentation is sufficiently resolved}. We conjecture that the corresponding criterion for Eulerian methods, which specify a spatial resolution $\Delta x$ (which may be either fixed or adaptive) is:
\be
\Delta x \lesssim 0.2 L_\mathrm{sonic} \approx 0.2 R_\mathrm{cloud}\,\mach^{-2},
\label{eq:convergence_crit_grid}
\ee
meaning that the sonic length $L_\mathrm{sonic}\approx R_\mathrm{cloud} \mach^{-2}$ is resolved across a certain number of cells. We expect the scaling $\propto \mach^{-2}$ to hold, but we caution that the exact numerical coefficient, encoding the exact number of cells required, may not generalize to other methods -- it will generally depend upon the specifics of the MHD and gravity solvers used. For AMR methods, Equation \ref{eq:convergence_crit_grid} may impose some requirement for both the refinement criterion {\em and} the base grid resolution; \citet{Haugbolle_Padoan_isot_IMF} found that it is necessary to scale both the base {\em and} maximum AMR resolution levels to achieve convergence.


\begin{figure}
\begin {center}
\includegraphics[width=0.95\linewidth]{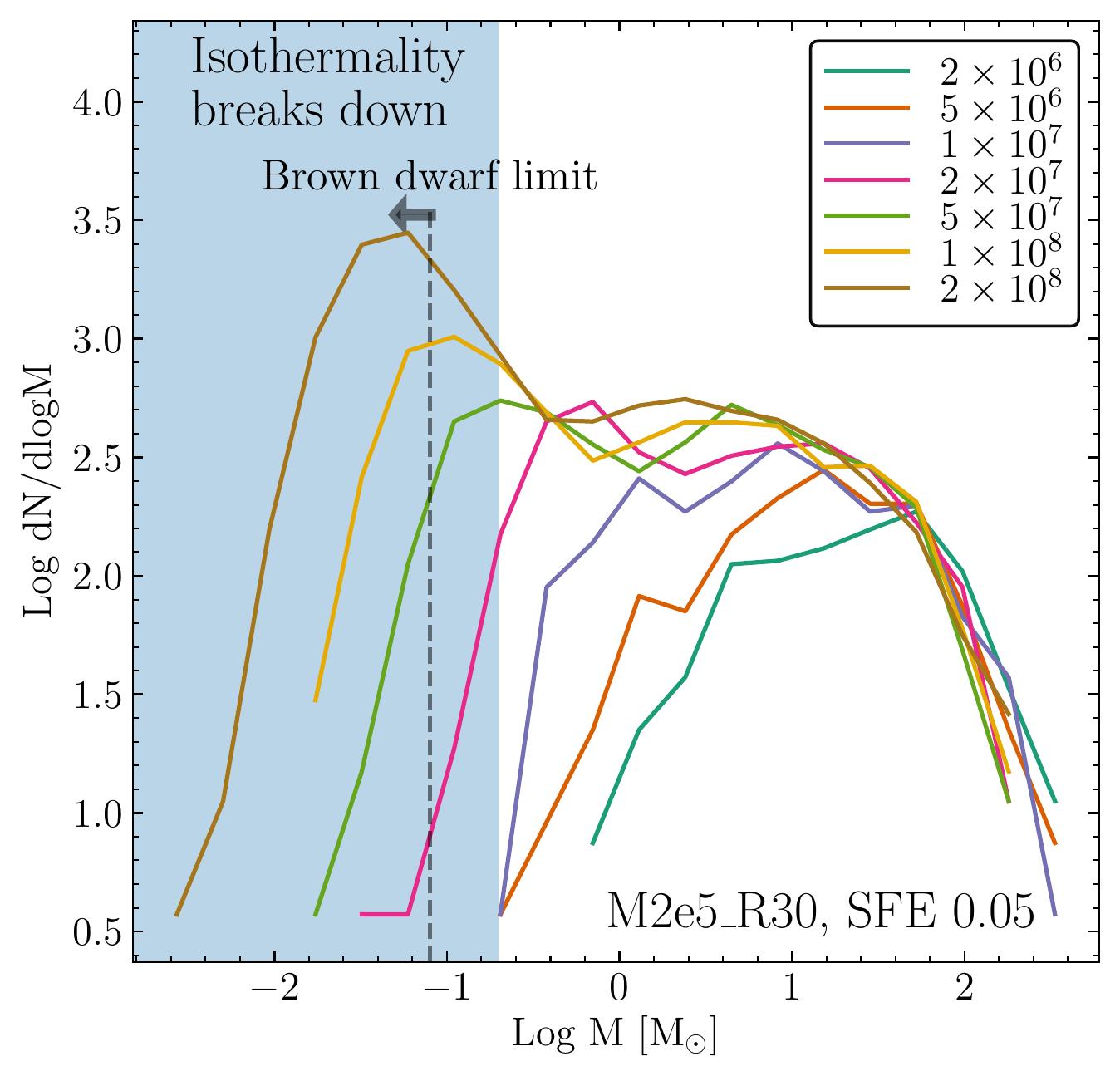}
\vspace{-0.4cm}
\caption{Mass PDF of sink particles at 5\% SFE for the \textbf{M2e5\_R30} run ($M_0=2\times10^5\,\msun$, $\alphaturbzero=2$, $\mach=29$, see Table \ref{tab:IC}) at various mass resolutions ($M_0/\Delta m$). Note that unlike Figure \ref{fig:stat_examples}, here we plot the full range of sink particle masses. We shaded the region where non-isothermal effects are expected to suppress the formation of new sinks \citep{bate_2009_rad_importance, Offner_2009_radiative_sim, Lee_Hennebelle_2018_EOS} and mark the brown dwarf regime ($M<0.08\,\msun$, dashed line). While the high mass end (that contains most of the mass) is insensitive to resolution (see Figure \ref{fig:percentiles}), a resolution sensitive peak forms near the resolution limit for $M_0/\Delta m\rightarrow \infty$, leading to an excess of brown dwarfs. Note that this low-mass peak is purely numerical, see Appendix \ref{sec:erratum}.}
\label{fig:low_mass_end}
\vspace{-0.5cm}
\end {center}
\end{figure}

Note that while it only contains a small fraction of the total IMF mass in these simulations, the {\em low mass} end of the IMF is clearly {\em not} converged and depends strongly on resolution in our simulations. Plotting the full IMF as a function of resolution in Figure \ref{fig:low_mass_end} we see that the \myquote{brown dwarf excess} predicted by ideal MHD physics alone becomes more severe as our resolution increases (note that this effect is numerical, see Appendix \ref{sec:erratum}). So we emphasize that our conclusions about $M_{50}$ and resolution-independence apply {\em only} to the relatively {\em large} masses containing most of the mass in the IMFs here. Note that it is unclear if this would still be true at much higher mass resolutions ($M_0/\Delta m\sim 10^{10}$), but probing that regime is prohibitively expensive with our current code. We find that this large number of very low mass sinks originate from dense regions around massive stars. Note that in these regions our assumption of isothermality is expected to break down, preventing further fragmentation in the gas and the formation of this \myquote{brown dwarf excess} (for discussion see \S~\ref{sec:tidal}). Furthermore, we find that this region of the IMF is sensitive to the details of our angular momentum return algorithm, but the conclusions of our study is not.

\section{Discussion}

\subsection{Comparison with other simulation studies}

There have been several studies in recent years that investigated the sink particle mass spectrum in simulations including MHD turbulence and gravity. In Table \ref{tab:comparison} we apply our fitting functions from Equations \ref{eq:fitting_func} and \ref{eq:mmax_fit} to the initial conditions of their simulations and compare them with the mass-weighted median and maximum sink mass in their reported IMFs. \citet{Haugbolle_Padoan_isot_IMF}, \citet{Lee_2019_binary_MHD}, and \citet{Federrath_2017_IMF_converge_proceedings} all used a simulation setup essentially identical to our {\it Box} simulation suite, simulating isothermal MHD with gravity and sink particles with the {\small RAMSES}, {\small ORION2}, and {\small FLASH} codes respectively. Compared to ours, these studies have subtle differences in the details of turbulence driving, but our results suggest these are unlikely to strongly affect the IMF (Figure \ref{fig:box_vs_sphere}). 

First, we compare with \citet{Haugbolle_Padoan_isot_IMF}. Most of these simulations included a prescription to model protostellar outflows, by having sink particles accrete only half of the inflowing mass and delete the rest, so we compare with the IMF from their {\it acc} test run that does have this prescription (their Fig. 14). We find that our predicted $\Mmedian=7.5 M_\odot $ and $M_\mathrm{\star,max}=19$ are quite close to their values of $4.2 \msun$ and $17 \msun$, both $<2 \sigma$ compatible if we estimate errors by bootstrapping their mass distribution and taking the RMS error of our fit.  We find even better agreement with the values in \citet{Lee_2019_binary_MHD}. 

Our prediction for $M_\mathrm{\star,max}$ matches the results of the {\it HighResIso} simulation in \citet{Federrath_2017_IMF_converge_proceedings}, but for those initial conditions we predict $\Mmedian = 11 M_\odot$, much greater than their $\Mmedian = 1.9 M_\odot$. This simulation produced 23 objects of mass $>1M_\odot$, so while the sampling of the IMF is certainly sparse, the numbers are not so small that we can readily attribute a factor of $\sim  5$ discrepancy to statistical variations. One difference between our respective calculations is that they used a mixture of compressive and solenoidal driving, vs. the purely solenoidal driving used in our {\it BOX} simulations. However given the robustness of our results to the details of turbulent forcing, this is unlikely to strongly affect the result either. We are left with no clear explanation for the discrepancy.

\citet{Wurster_2019_no_magnetic_break_catastrophe} simulated a $50 M_\odot$ dense clump akin to our {\it Sphere} suite, with both ideal and non-ideal smoothed-particle radiation MHD; we compare with their $\mu=5$, ideal MHD model, but note that they found that the IMF is not strongly affected by $\mu$ or non-ideal MHD effects. Our predictions of $\Mmedian \sim M_\mathrm{\star,max} \sim 1 M_\odot$ agrees very well with their results. As such, while it has been shown that accounting for full radiation transfer is important for suppressing brown dwarf formation \citep{bate_2009_rad_importance, Offner_2009_radiative_sim}, isothermal MHD may be a sufficient approximation to predict $\Mmedian$ and $M_\mathrm{\star,max}$.

Finally, we compare with \citet{padoan_2019_massive_sf}, who ran a $250\mathrm{pc}$ {\it Box}-type setup containing $1.9\times10^6\,M_\odot$, but with turbulence driven by supernova explosions. We derive approximate RMS $\mach$ and $\alphaturbzero$ values of 66 and 4.7 respectively, from the energy statistics given in \citet{Padoan_2016_sne_driving}, however we emphasize that these are rough values because 1. their ISM is not isothermal but rather multi-phase and 2. the energetics are highly variable and 3. the results in \citet{padoan_2019_massive_sf} are from a different, higher-resolution simulation with the same physical parameters. Nevertheless we predict $\Mmedian = 36 M_\odot$, within a factor of 2 of their value of $\sim 20M_\odot$. They attribute this overprediction of the IMF turnover to a lack of numerical resolution, but our results suggest that they are actually close to the ``converged" value. Rather, we believe other, important processes that shape the IMF were neglected, as we will argue further in this section. 

In summary, we find that our simulations predict $\Mmedian$ and $M_\mathrm{\star,max}$ in very good agreement with the predictions of other codes running similar problems, with the exception perhaps of the {\small FLASH} simulations in \citet{Federrath_2017_IMF_converge_proceedings}. Whether this represents any meaningful difference in code behaviours, or sensitivity to prescriptions, can ultimately only be answered by a controlled code comparison study \citep[e.g.][]{Federrath_2010_sinks}. Overall the good agreement between the present study, \citet{Haugbolle_Padoan_isot_IMF}, \citet{Lee_2019_binary_MHD}, \citet{Wurster_2019_no_magnetic_break_catastrophe}, and arguably \citet{padoan_2019_massive_sf} is encouraging, suggesting that these IMF predictions have some robustness to choice of MHD solver and numerical sink particle prescriptions.



\begin{table*}
    \setlength\tabcolsep{3.0pt} 
	\centering
		\begin{tabular}{|c|c|c|c|c|c|c|c|c|}
		\hline
		Study & $M_0$ & $\mach$ & $\alphaturbzero$ & SFE [\%] & $\Mmedian$ (sim.) & $\Mmedian$ (Eq. \ref{eq:fitting_func}) & $M_\mathrm{max}$ (sim.) & $M_\mathrm{max}$ (Eq. \ref{eq:mmax_fit})\\
		\hline
		\citet{Federrath_2017_IMF_converge_proceedings} & 775 & 5 & 0.62 & 10 & 1.9 & 11 & 15 & 13 \\	
        \citet{Haugbolle_Padoan_isot_IMF} & 3000 & 10 & 1 & 10.8 & 4.2 & 7.5 & 17 & 19 \\ 
		\citet{Lee_2019_binary_MHD} & 601 & 6.6 & 1.2 & 6.6 & 6.7 & 6.8 & 12 & 7.6\\ 
		\citet{Wurster_2019_no_magnetic_break_catastrophe} & 50 & 6.4 & 2 & 15.2 & 0.9 & 1.3 & 1.2 & 1.2 \\ 
		\citet{padoan_2019_massive_sf} & 1.9e6 & 66 & 4.7 & 1.2 & 20 & 36 & 130 & 290 \\ 
		\hline
		\end{tabular}
        \vspace{-0.1cm}
 \caption{IMF results from previous simulations of MHD star formation in the literature. We compare the values of $\Mmedian$ and $M_\mathrm{\star,max}$ from the published IMFs to the prediction from the fits to our simulation results (Eqs. \ref{eq:fitting_func} and \ref{eq:mmax_fit}). All masses are given in $M_\odot$. All virial parameters are converted to the definition used in this work, $\alpha_\mathrm{turb}=5 \sigma_\mathrm{v}^2 R_\mathrm{cloud}/3 G M_\mathrm{0}$, using the volume-equivalent $R_\mathrm{cloud}\approx 0.6L_\mathrm{box}$ for box simulations. In studies that survey $\mu$, we compare with the one that is closest to our fiducial 4.2, however we do not expect varying $\mu$ to strongly affect results (\S\ref{sec:mu_insensitivity}).} 
 \label{tab:comparison}\vspace{-0.5cm}
\end{table*}

\subsection{Can ideal MHD alone explain the observed IMF?}

By transforming our results back to a dimensional form we can examine whether isothermal, ideal MHD and gravity alone are enough to explain the observed stellar IMF, as proposed by studies such as that of \citet{Haugbolle_Padoan_isot_IMF}. At first our results might seem to support this conclusion, as we show that such a system forms stars with a well-defined, resolution-insensitive characteristic mass, which corresponds to a ``turnover mass'' in the IMF: above this mass the predicted mass spectrum is close to the observed power law of \citet{salpeter_slope}, while below that value it becomes shallower, like the observed IMF \citep{imf_review}. 

However, there are three major discrepancies between this predicted behavior and the observed IMF: (1) the predicted characteristic mass is much too large, for typical cloud conditions; (2) the characteristic mass depends sensitively on cloud properties, predicting far too much scatter in IMFs across different star-forming regions; (3) the low-mass end of the IMF has the wrong slope, and predicts an excess of brown dwarfs which is progressively more severe at higher resolution (with a shape that is dependent on the specific numerical implementation).

\begin{figure*}
\begin {center}
\includegraphics[width=0.48\linewidth]{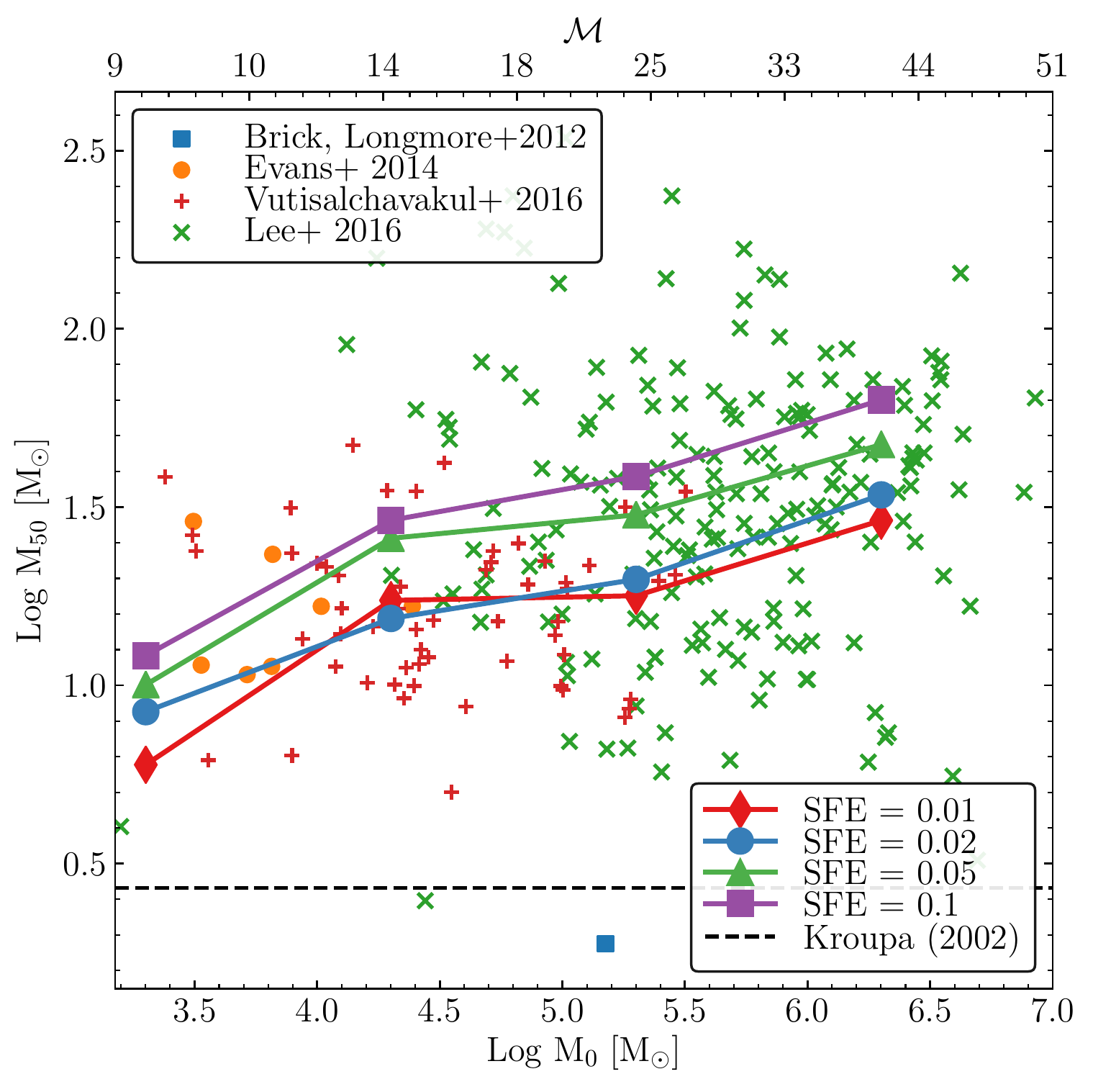}
\includegraphics[width=0.48\linewidth]{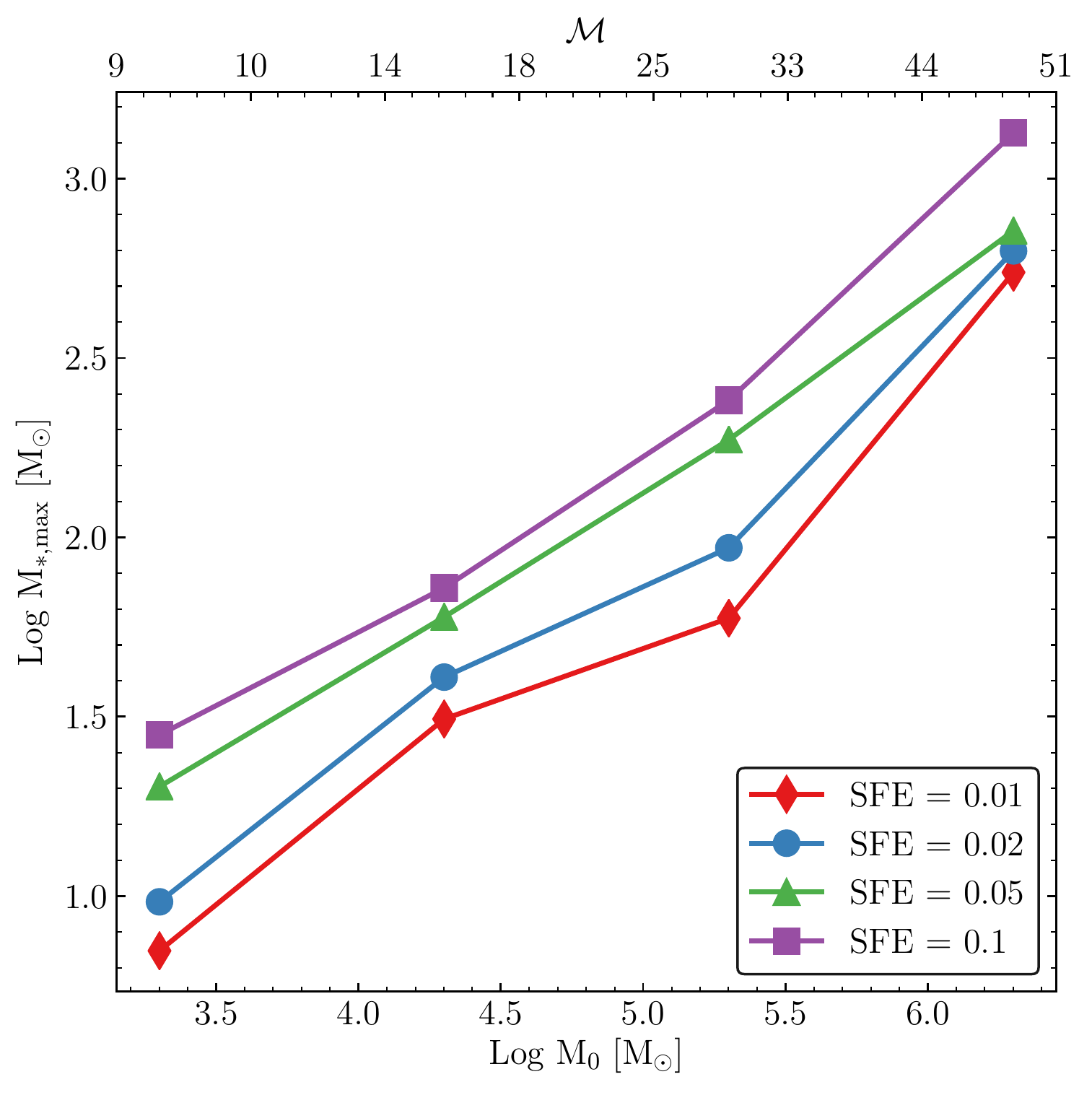}
\vspace{-0.4cm}
\caption{The mass-weighted median sink mass $\Mmedian$ (left) and the maximum stellar mass $M_{\star,\mathrm{max}}$ (right) as a function of initial cloud mass $M_{0}$ at different star formation efficiencies (labeled), for clouds chosen from Table \ref{tab:IC} to have the {\em same}, single virial parameter $\alphaturbzero=2$, same sound speed $\cs=0.2\,{\rm km\,s^{-1}}$ and lie {\em exactly} on the local Solar-neighbourhood median mass-size and linewidth-size relation of GMCs (corresponding Mach number $\mach$, since this is one-to-one with mass for this restricted simulation set, is shown in the top axis). The dashed vertical line on the left denotes $\Mmedian$ for a fit to the observed IMF using the \citet{kroupa_imf} form. The symbols show the predicted $\Mmedian$ values by Equation \ref{eq:m50_scaling_MW} using observed properties of nearby molecular clouds \citep{Evans_2014_MW_GMC_SFR}, MW GMCs \citep{Lee_eve_2016_GMC_SFE, Vutisalchavakul_2016_MW_GMC_SFR} and the extremely dense `Brick' cloud in the Galactic centre \citep{Longmore_2012_the_Brick}. Note that we assumed a constant $\cs=0.2\,{\rm km\,s^{-1}}$ for observed clouds, which likely reduces the scatter in the results by a significant amount. We find that the predicted IMF masses are (1) order-of-magnitude larger than observed (with $\sim 1000\,M_{\odot}$ stars routinely forming in massive clouds), (2) depend significantly on time (SFE), and (3) depend strongly on cloud properties. Due to the scatter in $\alpha_{\rm turb}$ and deviations from the mass-size and linewidth-size relations, for observed clouds this model predicts significant scatter and wildly varying IMFs for more extreme environments (e.g., the Brick).}
\label{fig:M50_GMC_mass}
\vspace{-0.5cm}
\end {center}
\end{figure*}

First, consider (1) in more detail. We find that, for conditions similar to a typical MW GMC, the simulations predict an IMF turnover of $\sim 20 M_{\odot}$ (see Figure \ref{fig:M50_GMC_mass}). Meanwhile, using the \citet{kroupa_imf} form for the observed IMF with an appropriate high mass cut-off ($200\,\msun$) we get $\Mmedian$ of $\sim 2\,\msun$, an order of magnitude lower than predicted by our model. Even if we account for feedback (e.g., winds, jets) reducing accretion by applying a correction factor of 2-3 (similar to \citealt{Haugbolle_Padoan_isot_IMF}) the predicted characteristic mass still ends up a factor of 3-5 larger than that observed. One might worry that this is because massive stars are allowed to accrete, in principle, for longer than their main sequence lifetimes (since we ignore any stellar evolution), but we find that even if we ``delete'' massive sinks after their main sequence lifetimes this has very little effect on our results, owing to fast and efficient new sink formation in the simulated GMCs (and the fact that most of the accretion onto these sinks occurs very quickly after they form; see Figures \ref{fig:starforge_series} and \ref{fig:sfe_t_evol}). One more thing to note is that our highest-resolution simulations reach an effective Jeans-length resolution of  $\sim<1\,\mathrm{AU}$ (in the case of {\bf M2e3\_R3} at maximum resolution), so unresolved binary formation is unlikely to significantly decrease our sink masses. Even if we took the extreme case and compared the predicted $\Mmedian$ with that of the system IMF \citep{chabrier_imf}, it would only account for a factor of $\sim 2$ shift.

We can also see that our predicted stellar masses are too large by considering the masses of the most massive stars forming in typical clouds. We find that in massive GMCs (total complex mass $\sim 10^6\,\msun$) stars with $\sim 1000\,\msun$ masses form routinely in the simulations (Figure \ref{fig:M50_GMC_mass}). These are far more massive than the most massive stars seen in current observations \citep{Crowther_most_massive_stars_R136_2016}, although admittedly if such stars do exist their lifetimes would be extremely short.

Regarding point (2), another significant issue is the dependence of $\Mmedian$ on the initial conditions of the cloud. While we find our results to be insensitive to some details of the ICs (e.g., driven vs decaying turbulence) $M_{50}$ is sensitive to the initial cloud mass $M_{0}$, sonic Mach number $\mach$, turbulent virial parameter $\alphaturbzero$, and star formation efficiency (SFE), according to Eq.~\ref{eq:fitting_func}. Observed clouds exhibit an order of magnitude scatter in observed virial parameter \citep{kauffmann_pillai_2013, heyer_dame_2015}, which would translate into a similar ($\sim 1\,$dex) cloud-to-cloud scatter in $\Mmedian$, in the simulations here. Even assuming that all GMCs have a constant $\alphaturbzero=0.1$ (the required value to have $\Mmedian\approx M_{\mathrm{obs}}$, even though the observed average is closer to  $\alphaturbzero\sim 2-3$, see \citealt{heyer_dame_2015, Miville_Deschenes_2017_MW_GMCs}), in dimensional units this would mean $M_{50} \propto {\rm SFE}^{0.3}\,M_{0}^{0.2}\,\Sigma^{-0.8}\,\cs^{3.2}$ (Eq.~\ref{eq:m50_scaling_MW}). Observed instantaneous cloud SFEs ($M_{\ast}/M_{0}$) in nearby well-studied GMCs vary by 3 orders of magnitude ($\sim1\,$dex 1-$\sigma$ scatter; see e.g., \citealt{eve_lee_2016_GMC_sfe}), predicting $\sim 1$\,dex spread in the characteristic IMF masses of these nearby clusters. Even if {\em this} was fixed, the result is extremely sensitive to the cloud temperature, which varies by factors of several, again predicting $\sim 1$\,dex spread in $M_{50}$. It should be noted that some of these properties co-vary following e.g. the linewidth-size or size-mass relations. In Figure \ref{fig:M50_GMC_mass} we plugged observationally inferred properties of MW clouds from various catalogs into Equation \ref{eq:m50_scaling_MW} and found about a dex of scatter in $\Mmedian$. It should be noted that different catalogs utilize different methodologies (see \citealt{grudic_2018_mwg_gmc} for a summary), including different tracers for gas (dust vs CO) and stellar mass (free-free emission vs IR vs YSO counts), which, combined with the uncertainties of other observationally inferred properties like the cloud virial parameter, leads to order of magnitude uncertainties in the predicted $\Mmedian$. Nevertheless, by looking at more extreme regions, like the Central Molecular Zone of the MW, starburst or high redshift galaxies, we find surface densities a factor $100-1000$ higher than in the MW \citep{Solomon_1997_ULIRG_ISM, Swinbank_2011_dense_galaxy_ISM}, predicting drastically more bottom-heavy IMFs than in the MW, since $M_{50}\propto \Sigma^{-0.8}$ (see the `Brick' in Figure \ref{fig:M50_GMC_mass}). In short, as shown in detail in \citet{guszejnov_imf_var, guszejnov_extragal_imf}, a scaling of $M_{50}$ with cloud properties of the sort predicted here would predict order-of-magnitude variation in the IMF turnover mass in the Milky Way Solar neighbourhood and more in nearby galaxies, contrary to the observed near-universality of the IMF in the local Universe  \citep{imf_review,imf_universality}. 

Finally, (3): as discussed above, at low (sub-Solar) masses the IMF predicted by ideal MHD does not exhibit any converged turnover down to the smallest resolved masses in our simulations (sub-Jupiter). In fact the IMF {\em steepens} progressively at very low masses, predicting even more sub-stellar objects, every time we increase our resolution. So there is a clear discrepancy with observations (excess of brown dwarfs and smaller objects), and ideal MHD cannot robustly predict the IMF shape in this regime.

These conflicts with observations indicate that \emph{isothermal, ideal MHD with gravity and no additional physics cannot explain the observed IMF}.

It should also be noted that star formation in the simulation proceeds very efficiently, reaching 10\% SFE in one freefall time ($\epsilon_{\mathrm{ff}}\sim 0.1$, see Figure \ref{fig:sfe_t_evol}), and continues (at an accelerating pace) until an order unity fraction of the gas is turned into stars. Meanwhile, observations indicate that typical GMCs convert only a few \% of their mass into stars by the end of their lifetimes \citep[see e.g.,][]{sf_big_problems}. This is yet another obvious indication that the physics here is incomplete.

We should also note that while ideal MHD does appear to predict a plausible Salpeter-like {\em slope} for the massive end of the IMF, this is not a unique effect of ideal MHD, but in fact emerges just as robustly in isothermal non-MHD simulations \citep{guszejnov_isothermal_collapse}, as a generic consequence of turbulent fragmentation \citep{general_turbulent_fragment}, competitive accretion \citep{bonnell_2007_competitive_accretion_imf}, or indeed {\em any} process which is scale-free over a sufficient dynamic range \citep{guszejnov_scaling_laws}.

\subsection{Potential roles for additional physics in setting the IMF}


\subsubsection{The opacity limit and tidal forces}\label{sec:tidal}

Isothermality is a key assumption in the current simulations. But even at low densities, it is debatable whether this is a good assumption, and it must break down at the highest densities where protostars form. Recent works have revived the idea of this transition (i.e. the traditional opacity limit) being responsible for setting the IMF (for the original idea see \citealt{lowlyndenbell1976, rees1976}) by taking into account the tidal screening effect around the first Larson core \citep{Lee_Hennebelle_2018_EOS,Colman_Teyssier_2019_tidal_screening}. These simulations mostly concentrate on the non-magnetized case, but \citet{Lee_Hennebelle_2019_T_B} investigated the inclusion of ideal MHD when including an idealized barotropic equation of state (meant to represent suppression of cooling above some limit) and claimed that the IMF characteristic mass is still set by the mass of the first Larson core ($M_{\mathrm{Larson}}\sim 0.02\,\msun$, leading to $M_\mathrm{tidal}\sim 0.2\,\msun$).

The simulations of \citet{Lee_Hennebelle_2019_T_B} were run on centrally condensed $1000\,\msun$ clouds with characteristic radius $0.084\,\pc$, $\mach\sim 22$, $\alphaturbzero \sim 1$, and $\mathrm{SFE} \sim 0.1$. Applying the $\Mmedian$ scaling from our results (Equation \ref{eq:m50_scaling}) leads to $\Mmedian\approx 0.1\,\msun$, comparable to the $0.1-0.2\,\msun$ peak coming from tidal screening around the first Larson core. So, in that case, the characteristic mass from isothermal MHD fragmentation happened to coincide with the mass scale imprinted by the Larson core, possibly explaining why introducing the magnetic field was not found to have a major effect. We showed in Figure \ref{fig:M50_GMC_mass} that, for initial conditions appropriate for MW GMCs, $\Mmedian\approx 20\,\msun$, much larger than this tidal screening mass. Since additional heating can only suppress fragmentation, we expect that adding the opacity limit to our calculation would imprint a low-mass cut-off scale upon the IMF, mitigating the brown dwarf excess and perhaps allowing the low-mass (sub-stellar) end of the IMF to exhibit robust numerical convergence. But the {\em high-mass} end of the IMF, including $M_{50}$ as studied here, lies far above this mass scale and would be unaffected (or even slightly {\em increased}) by accounting for inefficient cooling (and the tidal effects described above). 

In other words, \emph{tidal screening around the first Larson core should affect the IMF, but it alone is not sufficient to set the characteristic mass of stars.} Additional mechanisms are required to suppress the formation of massive stars.


\subsubsection{Non-ideal MHD terms}
Our assumption of ideal MHD is also expected to break down in the very dense gas within pre-stellar and protostellar cores and disks, in which the timescales for ambipolar diffusion, Ohmic resistivity, and the Hall effect can become comparable to the dynamical time. These effects may be important for preventing the magnetic braking that would otherwise prevent protostellar disks from existing (\citealt{hennebelle_fromang_2008_corecollapse, li_2011_mhd_braking, wurster_2016_nonideal_braking}, see however \citealt{Wurster_2019_no_magnetic_break_catastrophe} for a counterargument), determining the physical properties of disks \citep{hennebelle_2016_mhd_disks}. In the present work we have found that the dynamical effect of the magnetic field does play some role in inhibiting fragmentation, so in principle the breakdown of flux-freezing could permit smaller fragment masses. But the effect we see is weakly-dependent on magnetic field strength. Moreover, \citet{Wurster_2019_no_magnetic_break_catastrophe} investigated the combined effects of non-ideal MHD terms upon the IMF predicted by simulations and found no systematic difference compared to ideal MHD. And even if we imagined the ``most extreme non-ideal'' limit, where non-ideal terms allowed for either efficient de-coupling of magnetic fields from most of the gas (ambipolar diffusion) or efficient magnetic damping (resistivity), this would lead to results {\em more like non-MHD} simulations, which as discussed above fare {\em even more poorly} at predicting {\em any} IMF shape resembling that observed.

Based upon these arguments, we anticipate that the effects of non-ideal MHD upon the IMF itself are weak. Even if they are not weak, they cannot lead to the correct IMF shape.

\subsubsection{The necessity of feedback regulation}

While isothermal, ideal MHD does produce {\em an} IMF it has several issues as noted above: (1) too many massive stars, (2) sensitivity to cloud ICs, (3) too many brown dwarfs, and (4) excessive star formation continues until $\mathrm{SFE}\sim 1$ with very high star formation efficiency ($\epsilon_{\mathrm{ff}}\sim 0.1$). All of these, however, are likely to be strongly influenced by feedback processes that are ignored here.

Non-isothermal cooling physics is likely important for the excess of brown dwarfs (see \S~\ref{sec:tidal}). However, many authors have argued that it is also crucial to account for radiative heating by protostars as they accrete \citep{Offner_2009_radiative_sim, krumholz_stellar_mass_origin, bate12a, Myers_2013_ORION_radiation_IMF,guszejnov_gmc_imf,guszejnov_feedback_necessity}. Whether protostellar heating or other physics is the dominant physics at substellar mass scales remains to be fully explored, but such heating certainly has the desired qualitative effect of suppressing low-mass fragmentation.


In parallel, protostellar outflows and jets can expel a significant fraction (up to half or more) of the material accreted in a collapsing core, reducing the stellar masses directly \citep[e.g.,][]{Offner_Chaban_2017_jets_sfe}. These outflows can also drive turbulence on small scales \citep{Offner_Arce_2014,Offner_Chaban_2017_jets_sfe, murray_2018_jets} that can both disrupt the nearby accretion flow and drive the local region to form fragments with smaller characteristic masses (similar to increasing $\mach$ in our simulations). Thus protostellar outflows can, in principle, have a significant effect upon the IMF when included in simulations \citep{Cunningham_2011_outflow_sim, hansen_lowmass_sf_feedback, krumholz_2012_orion_sims, Federrath_2014_jets, Cunningham_2018_feedback}. They also tend to reduce the rate of star formation by modest factors ($\sim2-3$; \citealt{federrath_2015_inefficient_sf}), which would bring our SFE per-freefall-time ($\epsilon_{\mathrm{ff}}$) to a few percent. Thus protostellar outflows may be an important feedback mechanism that can regulate the star formation rate to observed levels, especially in regions where massive stars are absent \citep{grudic_2018_mwg_gmc, krumholz_2019_cluster_review}.

However, it is unlikely that protostellar outflows are powerful enough to regulate star formation on the scale of the entire GMC \citep{matzner_mckee_2000_jets, Murray_2010_GMC_disruption}. \emph{ Stellar} feedback, i.e., feedback mechanisms originating in main-sequence stars powered by nuclear fusion (including ionizing radiation, stellar winds, and supernova explosions), are likely responsible for regulating the integrated star formation efficiency of GMCs down to observed levels, by disrupting the cloud once sufficient stellar mass has formed (see \citealt{krumholz_2019_cluster_review} for review and Fig. 1  of \citealt{grudic_mond} for a literature compilation of theoretical predictions). For typical local GMCs, these mechanisms (given standard stellar evolution tracks) are more than sufficient to disrupt clouds after a few percent of the total mass is turned into stars \citep{grudic_2016, kim_2018_gmc_raytrace, Li_Vogelsberger_2019_GMC_disrupt}. This process must also affect the IMF, as it abruptly cuts off the gas supply for accretion, and could also potentially stir turbulence on small scales. \citet{Gavagnin_2017_SF_feedback} investigated the effect of photoionization feedback upon the IMF predicted in radiation-hydrodynamic simulations (neglecting magnetic fields), and found that it reduced the mean stellar mass of massive stars by a factor of $\sim 3$, from $\sim 15M_\odot$ to $\sim 5 M_\odot$. This is still more than an order of magnitude larger than the observed mean, so while ionizing radiation certainly has important effects in high-mass star formation (it is likely the dominant contributor to GMC disruption, see \citealt{grudic_2018_mwg_gmc}), it cannot account for the mass scale of the IMF alone.

Stellar winds can disrupt the gas around massive stars and prevent further accretion, thus potentially reducing the frequency of high mass stars, but their effects fall off quickly and have not been found to significantly affect either the IMF \citep{Dale_Bonnell_2008_winds_IMF} or the overall cloud SFE \citep{Dale_2013_wind_cloud_disruption} in simulations. However, to our knowledge no dynamical MHD star cluster formation simulations have investigated the effect of main-sequence stellar winds, and it is conceivable that magnetic fields could enhance their effect, suppressing the growth of instabilities and transporting momentum and energy beyond the extent of the wind bubble itself \citep[e.g.][]{offner_2018_mhd_feedback}.

Supernova explosions dominate the overall feedback momentum injected into the ISM \citep{leitherer_1999_starburst99}, and are generally agreed to be the most important feedback mechanism in galaxy formation \citep{hopkins2014_fire,Somerville_Dave_2015_galaxy_formation_review, naab_ostriker_galform_review, Hopkins_2018_sne_feedback, vogelsberger_galform_review}. However, their effect upon the IMF must be indirect, because they occur too late to significantly affect the evolution of dense clumps in which star clusters form. Their main role in star formation is likely maintaining the state of ISM turbulence on the scale of the galactic scale height and driving galactic outflows (via super-bubbles and chimneys), thus regulating the ISM gas densities and other ``environmental'' properties which set the properties of GMCs in turn  \citep[e.g.,][]{Hopkins_2011_SFR_self_regulate,hopkins_2012_galaxy_structure, Walch_2015_SILCC_ISM_SN,Padoan_2017_SN_driving_SFE,Seifried_2018_GMC_SN_driving, guszejnov_GMC_cosmic_evol}.


The processes discussed in this section and their effects on star formation will be investigated individually in the upcoming STARFORGE simulation suite (Guszejnov et al. 2020, in prep.).


\section{Conclusions}
 
We carried out a suite of high-resolution simulations of turbulent molecular clouds and showed that ideal, isothermal MHD does exhibit a characteristic mass scale ($M_{50}$) that is inherited by the mass distribution of collapsed objects (see Figures \ref{fig:stat_examples} and \ref{fig:percentiles}). This is in contrast to non-magnetized clouds, which exhibit no such scale. The characteristic mass appears to be set by the turbulent properties of the cloud as it (at any given time) only depends on the cloud mass, the initial sonic Mach number, virial parameter and the current star formation efficiency (see Eq. \ref{eq:fitting_func} and Figure \ref{fig:convergence}). We find that using different detailed initial conditions, with driven or decaying turbulence does not affect this result (see Figure \ref{fig:box_vs_sphere}).

The {\em shape} of the mass distribution of collapsed objects is qualitatively similar to the observed intermediate and high-mass IMF, as it reproduces a Salpeter-like slope with a turnover to a ``flat'' slope below this characteristic mass $M_{50}$ (see Figure \ref{fig:stat_examples}). However, this model of isothermal turbulence with ideal MHD and no additional physics has severe difficulties explaining the observed IMF because the predicted mass scale (1) is an order of magnitude larger than the observed IMF mass scale, (2) evolves strongly in time with the cloud star formation efficiency, and (3) sensitively depends on initial clouds conditions/properties in a manner that would predict order-of-magnitude cloud-to-cloud (and larger galaxy-to-galaxy) variation in the IMF mass scale. In addition, (4) isothermal MHD predicts an excess of brown dwarfs (no sub-stellar turnover), which becomes more severe at higher resolutions, and (5) the star formation efficiency is too large and rises rapidly until essentially all gas in GMCs is turned into stars. It is thus necessary to include the physics of proto-stellar and stellar feedback in addition to ideal MHD and gravity in any star formation theory that hopes to explain current observations, which we will show in detail in a future work (Guszejnov et al. 2020 in prep.).

\section*{Acknowledgements}
The authors thank Mark Krumholz and Philip Mocz for useful discussions, and Aaron Lee, James Wurster, Cristoph Federrath and Eve J. Lee for providing data for comparison. 
DG is supported by the Harlan J. Smith McDonald Observatory Postdoctoral Fellowship. MYG is supported by a CIERA Postdoctoral Fellowship.
Support for PFH was provided by NSF Collaborative Research Grants 1715847 \&\ 1911233, NSF CAREER grant 1455342, and NASA grants 80NSSC18K0562 \&\ JPL 1589742.
SSRO is supported by NSF Career Award AST-1650486 and by a Cottrell Scholar Award from the Research Corporation for Science Advancement. CAFG is supported by NSF through grants AST-1517491, AST-1715216, and CAREER award AST-1652522; by NASA through grant 17-ATP17-0067; and by a Cottrell Scholar Award from the Research Corporation for Science Advancement. 
This work used computational resources provided by XSEDE allocation AST-190018, the Frontera allocation FTA-Hopkins supported by NSF, 
NASA HEC allocation SMD-16-7592, and additional resources provided by the University of Texas at Austin and the Texas Advanced Computing Center (TACC; http://www.tacc.utexas.edu).



\bibliographystyle{mnras}
\bibliography{bibliography} 




\appendix

\section{Detailed scaling of $\Mmedian$ with cloud parameters}\label{sec:fitting}

In this appendix we examine in detail how the mass-weighted median sink mass $\Mmedian$ depends on the turbulent virial parameter $\alphaturbzero$, sonic Mach number $\mach$, normalized magnetic flux ratio $\mu$ and the star formation efficiency (SFE), and how well it is fit by Equation \ref{eq:fitting_func}.

\begin{figure}
    \centering
    \includegraphics[width=0.99\linewidth]{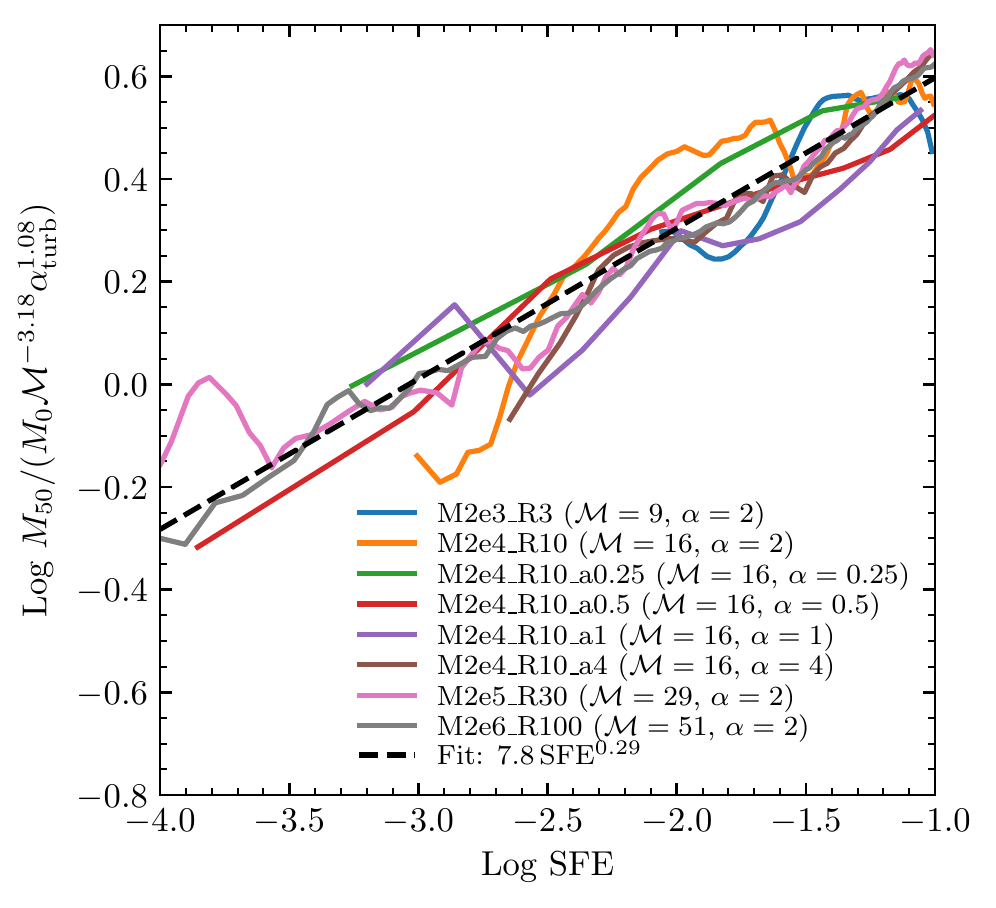}
    \vspace{-0.5cm}
    \caption{Comparison of the fit of $\Mmedian$ (Equation \ref{eq:fitting_func}) with simulation results, pulling out the best-fit scaling $\propto M_\mathrm{0}\mathcal{M}^{-3.18}\,\alphaturbzero^{1.08}$ and plotting as a function of SFE. The fit residuals have no clear trend with $\alphaturbzero$, $\mach$, or SFE, and tend to be smaller for higher-$\mach$ clouds that produce more sinks, indicating that fluctuations about the relation are statistical.}
    \label{fig:global_fit}
\end{figure}

Figure \ref{fig:global_fit} compares the fit from Equation \ref{eq:fitting_func} with the actual evolution of $\Mmedian$ in a subset of our runs which have various $M_0$, $\mach$ and $\alphaturbzero$ values. We find that all runs lie upon the predicted curve with deviations below 0.2 dex at all times and with no trend in the residuals with any of the input parameters, indicating that the fluctuations are likely statistical in nature.

\begin{figure}
\begin {center}
\includegraphics[width=0.95\linewidth]{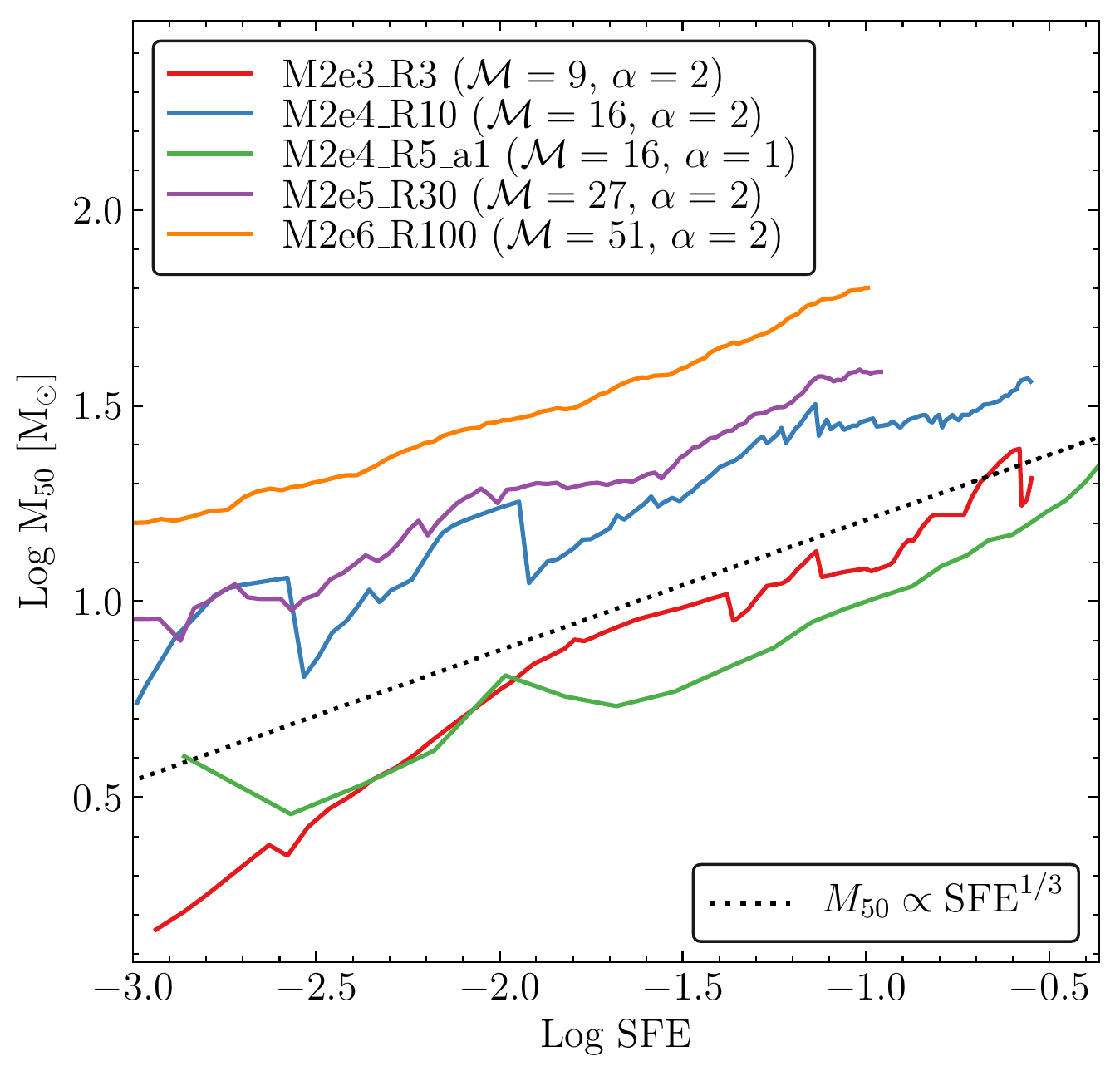}
\vspace{-0.4cm}
\caption{Evolution of the  mass-weighted median mass $\Mmedian$ as a function of star formation efficiency for a subset of runs. $\Mmedian$ increases with star formation efficiency in a power-law fashion, roughly consistent with $\Mmedian \propto \mathrm{SFE}^{1/3}$, regardless of the initial conditions.}
\label{fig:M50_sfe_evol}
\vspace{-0.5cm}
\end {center}
\end{figure}

\begin{figure*}
\begin {center}
\includegraphics[width=0.45\linewidth]{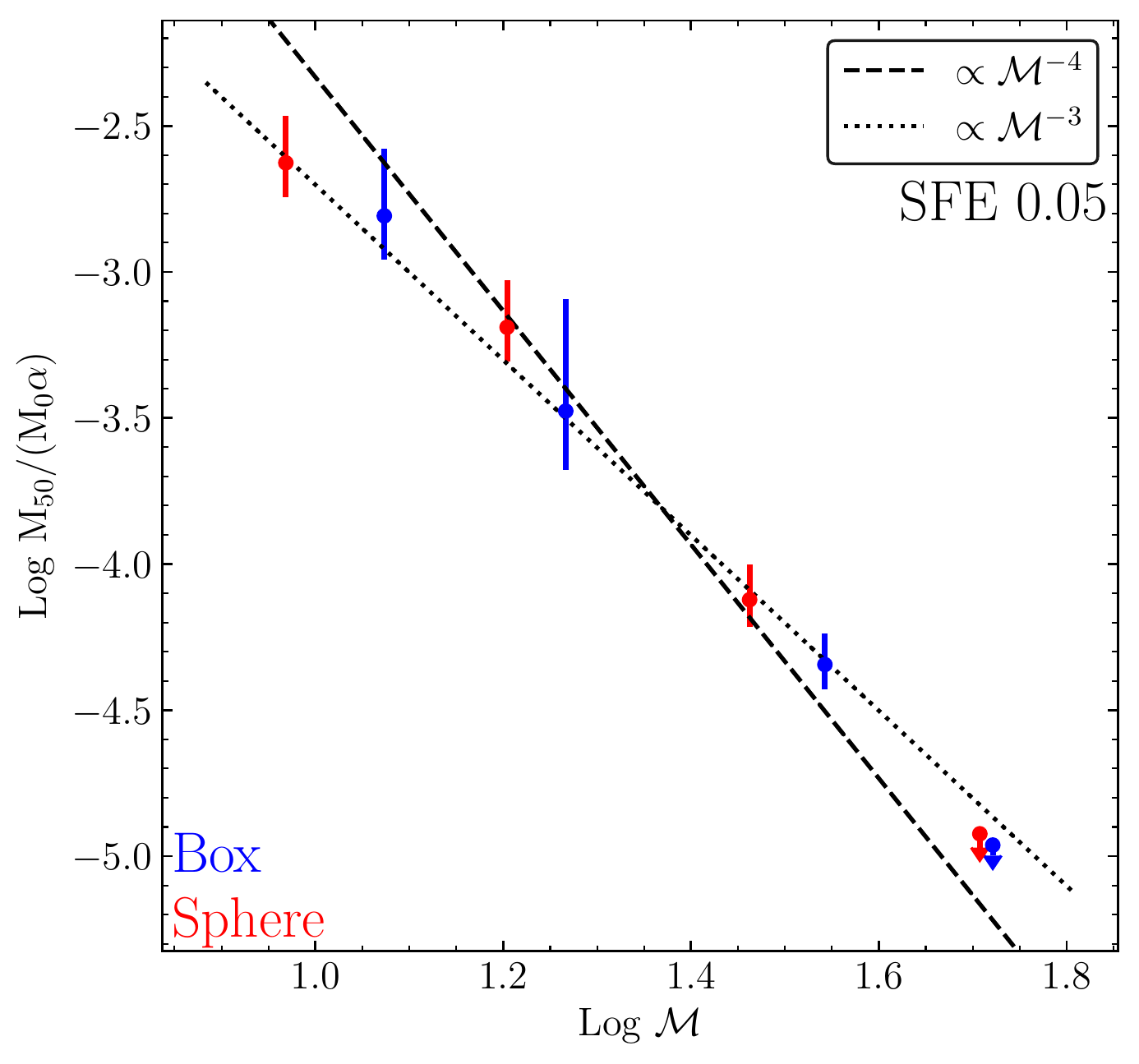}
\includegraphics[width=0.45\linewidth]{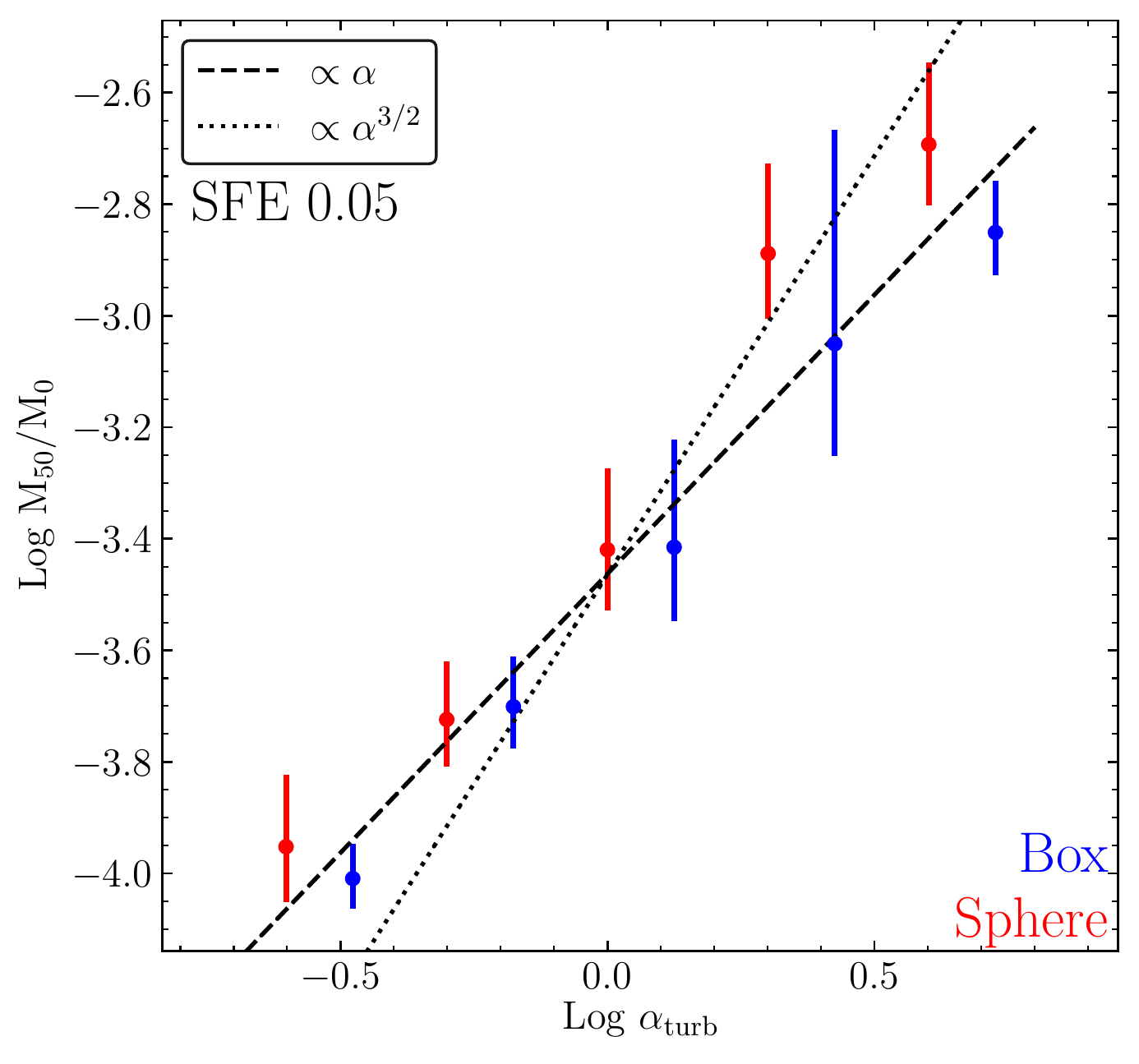}
\vspace{-0.4cm}
\caption{Dependence of the mass-weighted median sink mass $\Mmedian$ at 5\% SFE on the initial Mach number (\textit{left}) and turbulent virial parameter (\textit{right}). We show results for both \textit{Sphere} and driven \textit{Box} initial conditions (denoted with blue and red respectively). Note that due to the nature of the driving Box runs with different Mach numbers have slightly different virial parameters. To compensate for this in the top figure we use $\Mmedian/(M_0 \alpha)$. The errors are estimated by bootstrapping: we resample the sink mass distribution at fixed total stellar mass and calculate the 95\% confidence interval of $\Mmedian$ over these realizations, which we denote with errorbars. Note that the resolution of the highest Mach number and lowest virial parameter calculations do not satisfy Equation \ref{eq:convergence_crit}, so their mass-weighted medians should be considered upper limits and are denoted by arrows.} 
\label{fig:M50_dependence_a_M}
\vspace{-0.5cm}
\end {center}
\end{figure*}

To get a sense of the accuracy of the predicted exponents in Equation \ref{eq:fitting_func} we examine how $\Mmedian$ depends on each of them independently. Figure \ref{fig:M50_sfe_evol} shows that $\Mmedian$ evolves roughly as $\Mmedian\propto \mathrm{SFE}^{1/3}$ for all runs. Meanwhile, Figure \ref{fig:M50_dependence_a_M} shows how varying $\alphaturbzero$ and $\mach$ respectively changes $\Mmedian$ (for the effects of changing $\mu$ see Figure \ref{fig:M50_dependence_mu}). The scaling with virial parameter appears to be consistent with $\Mmedian/M_0\propto \alpha$ while the Mach number dependence is close to $\Mmedian/M_0\propto \mach^{-3}$.

To estimate the errors of the fitted exponents for $\alphaturbzero$ and $\mach$ we first estimate the errors in $\Mmedian$ using bootstrapping, which means resampling the sink mass distribution at fixed SFE and calculating the 95\% confidence interval of $\Mmedian$ over these new samples. Then we fit the exponents at our fiducial SFE (5\%) by using runs between which only a single parameter varies (see Figure \ref{fig:M50_dependence_a_M}). For the exponent of SFE we estimate its error by fitting a power-law to our different runs (in Figure \ref{fig:M50_sfe_evol}) and take the variance of the fitted values. We find the following fitting parameters and errors
\be
\Mmedian \propto  M_0\, \mach^{-3.24\pm 0.08}\, \alphaturbzero^{1.01\pm 0.09}\, \mathrm{SFE}^{0.30\pm 0.05}
\label{eq:1Dfitting_func}
\ee
Note that contrary to the fitting in Eq. \ref{eq:fitting_func} here we use only a subset of our runs and fit each slope individually (hence the slightly different exponents).

\begin{figure*}
\begin {center}
\includegraphics[width=0.33\linewidth]{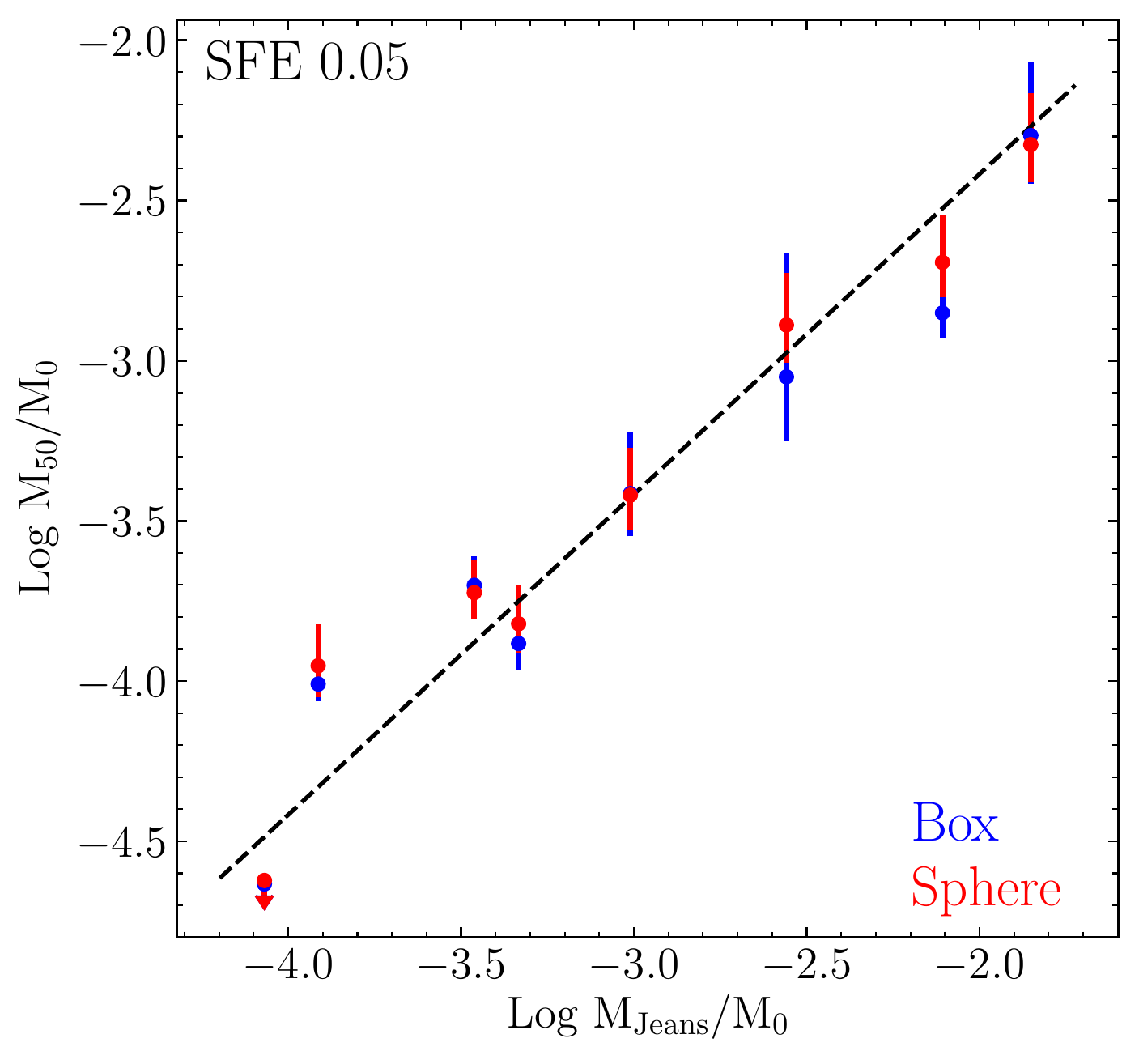}
\includegraphics[width=0.33\linewidth]{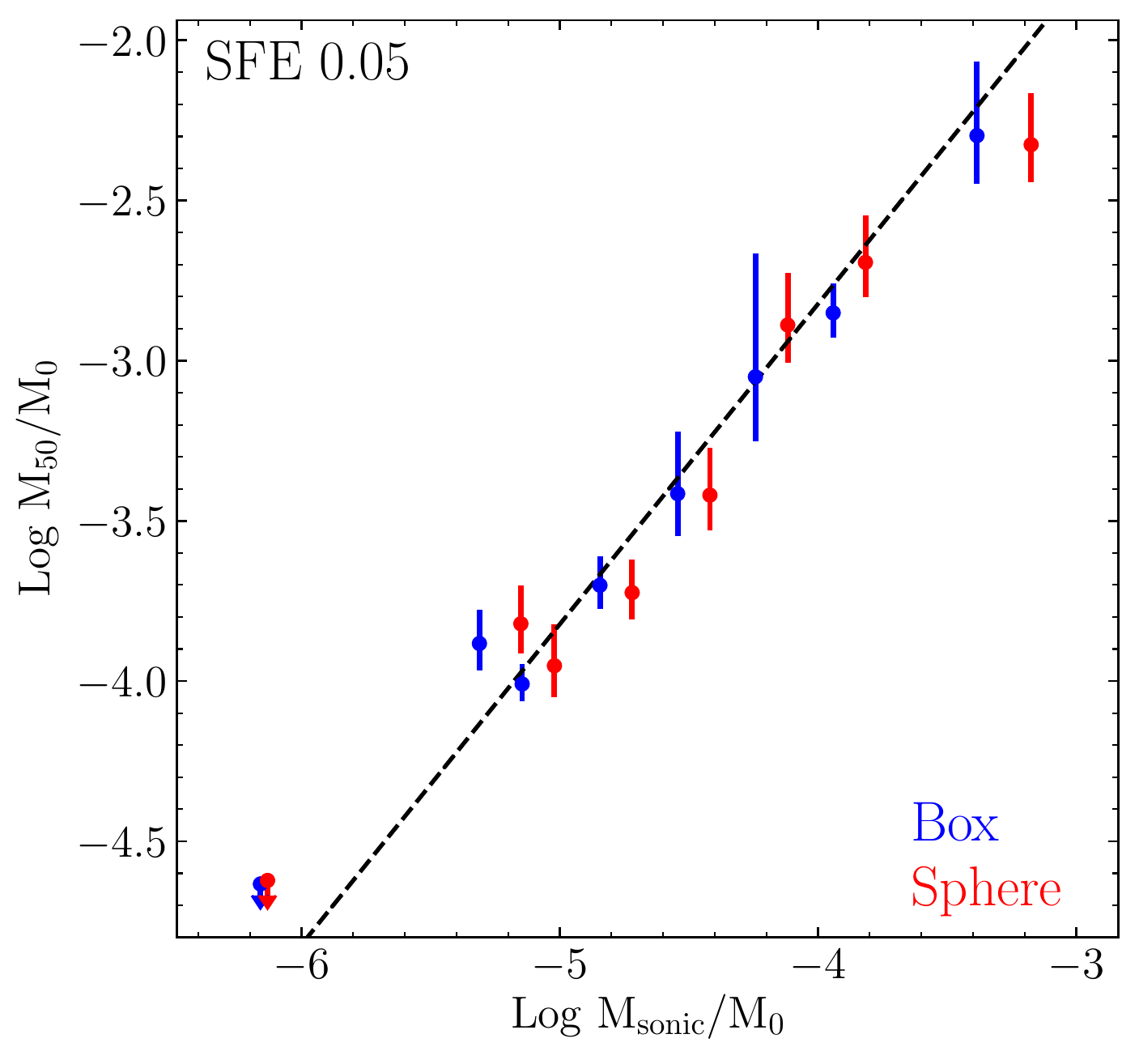}
\includegraphics[width=0.33\linewidth]{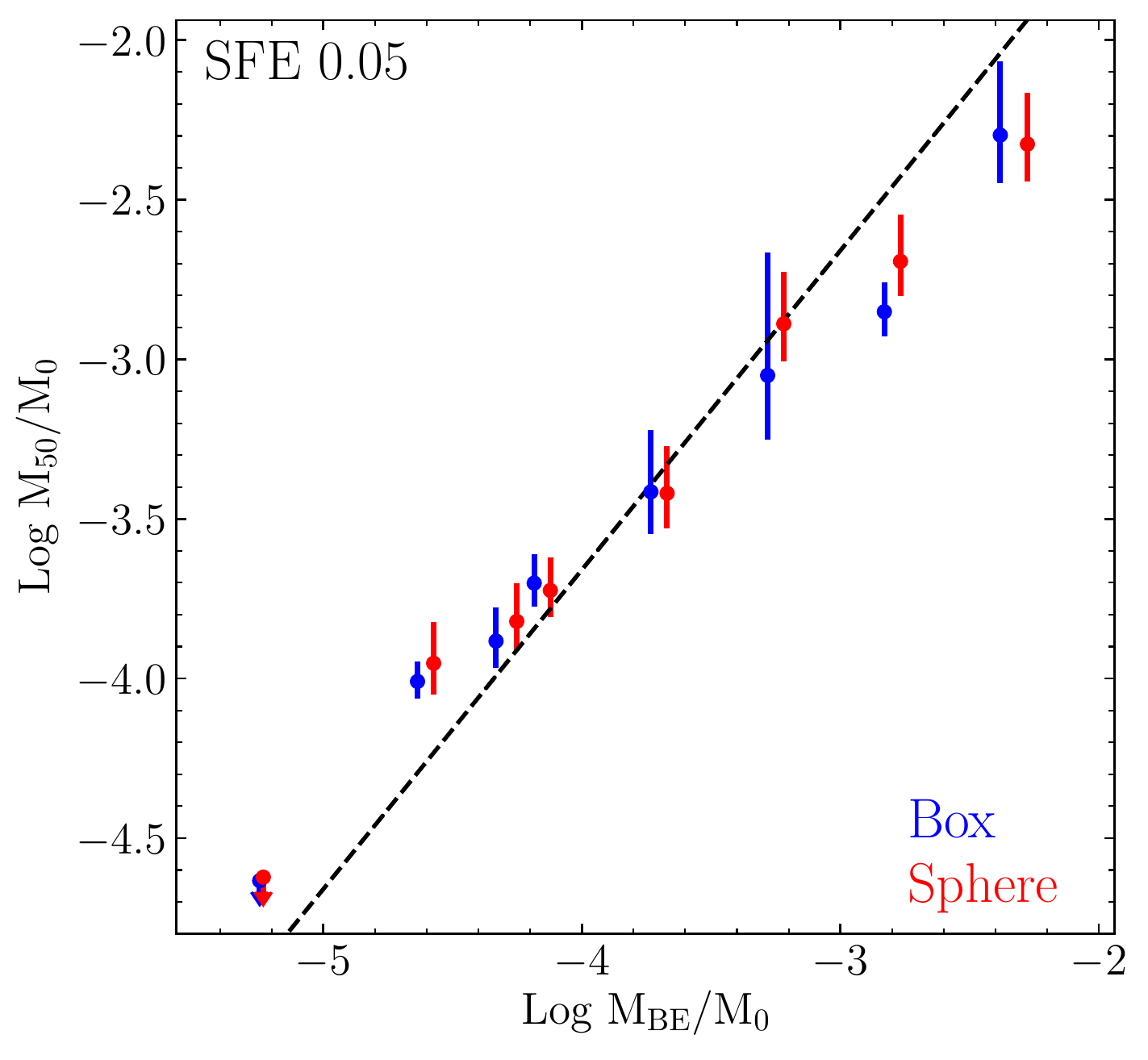}
\vspace{-0.4cm}
\caption{Dependence of the mass-weighted medan sink mass $\Mmedian$ on the Jeans mass $\MJeans$ (left, see Eq \ref{eq:mjeans_simple}), sonic mass $\Msonic$ (middle, see Eq \ref{eq:msonic_simple}) and the turbulent Bonnor-Ebert mass $\MBE$ (right, see Eq \ref{eq:MBE_simple}). Notation and errorbars are the same as in Figure \ref{fig:M50_dependence_a_M}.} 
\label{fig:M50_dependence_mass_scales}
\vspace{-0.5cm}
\end {center}
\end{figure*}

The exponents we find in Equations \ref{eq:fitting_func} and \ref{eq:1Dfitting_func} do not correspond to any of the known mass scales listed in \S~\ref{sec:params_scales} (see Equations \ref{eq:mjeans_simple}-\ref{eq:MBE_simple}). While neither mass scale is as good a fit as Equations \ref{eq:fitting_func} and \ref{eq:1Dfitting_func} (see Figure \ref{fig:M50_dependence_a_M}), Figure \ref{fig:M50_dependence_mass_scales} shows that they are all good qualitative predictors of $\Mmedian$ for our set of simulations.

\newpage

\section{Erratum}\label{sec:erratum}

\begin{figure}
	\begin {center}
	\includegraphics[width=0.95\linewidth]{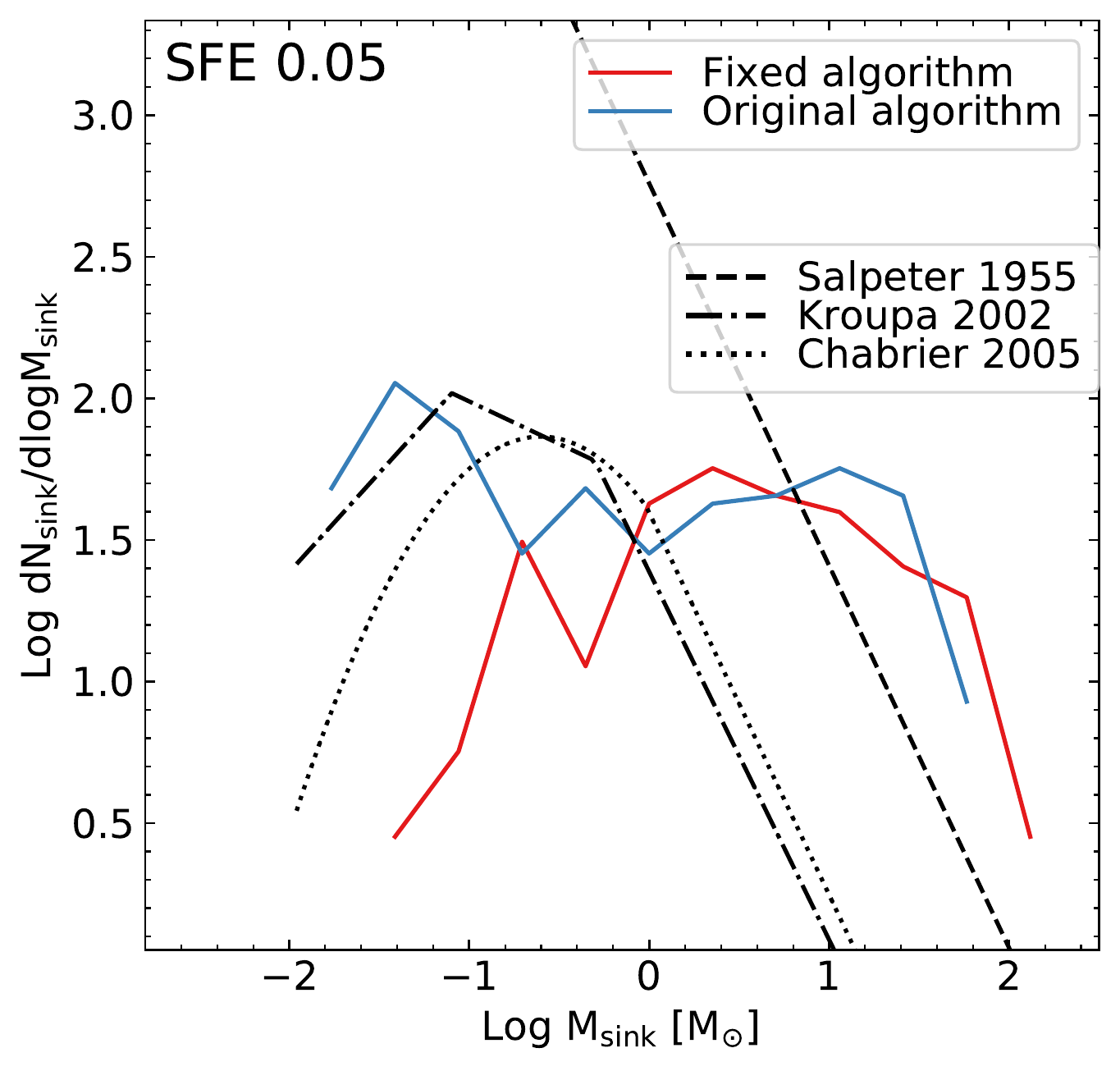}
	\vspace{-0.4cm}
	\caption{Distribution of sink particle masses measured in runs with both the original and the fixed algorithms for our \textbf{M2e4\_R10} initial conditions at 5\% star formation efficiency ($\mathrm{SFE}=\sum M_{\mathrm{sink}}/M_{0}$). We also show the \citet{salpeter_slope}, \citet{kroupa_imf} and \citet{chabrier_imf} fitting functions for the IMF. The peak at low masses with the original algorithm is clearly of numerical origin, however the high-mass ends in both cases are top-heavy compared to the observed one.}
	\label{fig:imf}
	\vspace{-0.5cm}
	\end {center}
\end{figure}

The paper \textit{Can magnetized turbulence set the mass scale of stars?} was published in MNRAS, 496, 5072-5088 (2020). In the original paper we found a large number of very low-mass sink particles (representing individual protostars) near the mass resolution limit (see Figure 10 of the original paper). After publication of the paper a detailed code review was carried out that found an uninitialized variable in the sink particle algorithm that  could occasionally lead to erroneous behavior. After re-running the simulation with the more thoroughly-developed sink particle methods used in \citet{starforge_methods} and \citet{starforge_jets_imf}, we found that this population of low-mass sink particles was drastically reduced (see Figure \ref{fig:imf}), suggesting that a sub-population of these was unphysical in origin (strengthening our conclusions about the necessity of additional physics to prevent an overly top-heavy IMF).

\begin{figure}
	\begin {center}
	\includegraphics[width=0.9\linewidth]{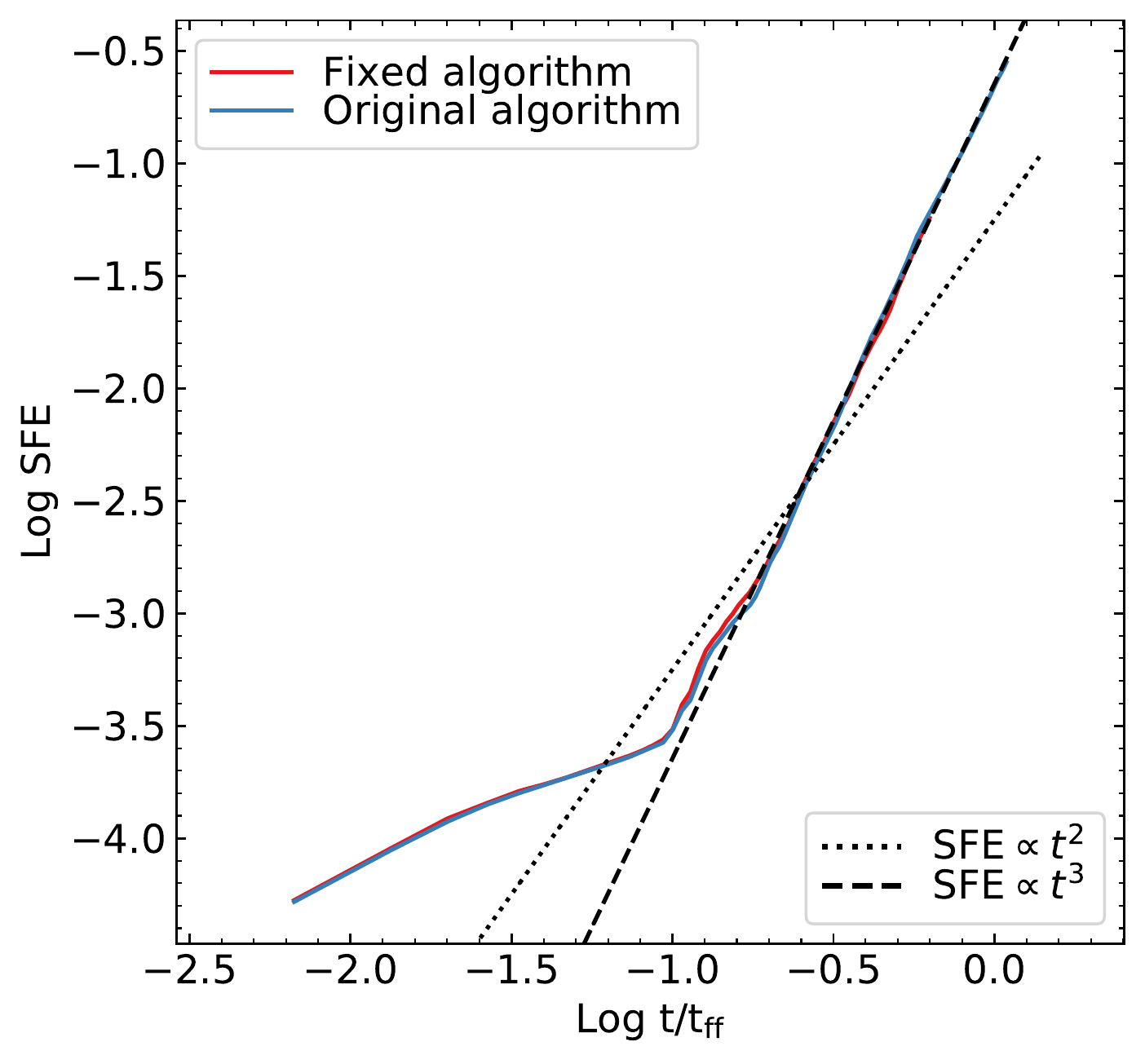}\\
	\includegraphics[width=0.9\linewidth]{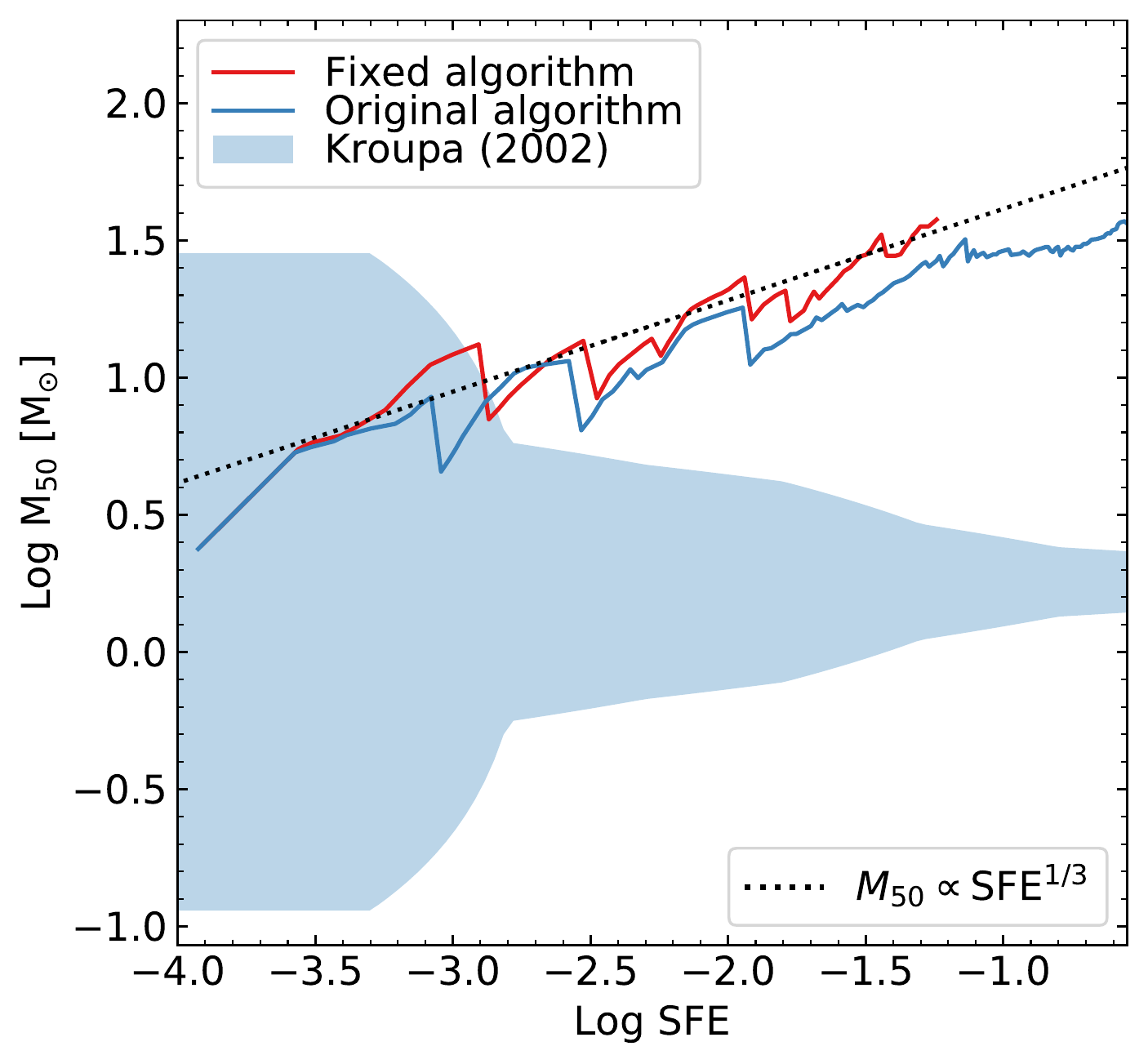}
	\vspace{-0.4cm}
	\caption{\textit{Left}:  Evolution of the star formation efficiency ($\mathrm{SFE}(t)=\sum{M_\mathrm{sink}(t)}/M_{0}$) as function of time with the original and the fixed version of the sink algorithm. Note that both versions of the code produce the same $\mathrm{SFE}\propto t^3$ behavior. \textit{Right}: The evolution of the mass-weighted median ($\Mmedian$, the mass scale above which half the total sink mass resides, right) sink mass as a function of star formation efficiency. We also show with a shaded region the 95\% confidence interval for these values if one sampled the \citet{kroupa_imf} IMF at the current SFE value in the cloud. The behavior with both the original and the fixed algorithms are essentially identical, leading to a top-heavy IMF.}
	\label{fig:evolplots}
	\vspace{-0.5cm}
	\end {center}
\end{figure}

Note that these low-mass objects represented a minor fraction of the total stellar mass. Since the main subject of our analysis was the mass-weighted median mass $\Mmedian$, the main conclusions of the original paper are not strongly affected by this issue, as shown by Figure \ref{fig:evolplots}. However we also conjectured that non-isothermal gas physics (e.g. the opacity limit for fragmentation) may be necessary to prevent an unphysically-large number of brown dwarfs from forming, as has been argued in many other works \citep{bate_2009_rad_importance,Offner_2009_radiative_sim,Lee_Hennebelle_2018_IC,Colman_Teyssier_2019_tidal_screening}. Because a significant number of the brown dwarfs predicted by the simulation were unphysical in origin, the actual factor by which the brown dwarf population must be suppressed was overstated, and potentially our assessment of the importance of additional physics in turn.

\bsp	
\label{lastpage}
\end{document}